\documentclass[aps,prl,reprint,groupedaddress]{revtex4-2}

\usepackage{glossaries}
\usepackage{graphicx}
\usepackage{xspace}
\usepackage{xcolor}
\usepackage{hyperref}
\hypersetup{
    colorlinks = true,
    linkcolor  = [rgb]{0.0,0.25,0.55},  
    citecolor  = [rgb]{0.0,0.35,0.65},  
    urlcolor   = [rgb]{0.0,0.3,0.6},    
}
\usepackage{cleveref}
\crefname{algocf}{alg.}{algs.}
\Crefname{algocf}{Algorithm}{Algorithms}

\usepackage{array}
\usepackage[ruled,vlined]{algorithm2e}

\usepackage{placeins}
\usepackage{booktabs}
\usepackage{multirow}

\newcolumntype{R}[1]{>{\raggedleft\arraybackslash}p{#1}}
\newcolumntype{C}[1]{>{\centering\arraybackslash}p{#1}}

\usepackage{orcidlink}
\usepackage{placeins}
\usepackage{booktabs}
\usepackage{amsmath}
\usepackage{amsfonts}

\newacronym{smc}{SMC}{Sequential Monte Carlo}
\newacronym{mcmc}{MCMC}{Markov Chain Monte Carlo}
\newacronym{ptmcmc}{PTMCMC}{Parallel Tempered Markov Chain Monte Carlo}
\newacronym{bbh}{BBH}{binary black hole}
\newacronym{bns}{BNS}{binary neutron star}
\newacronym{jsd}{JSD}{Jensen–Shannon divergence}
\newacronym{cbc}{CBC}{compact binary coalescence}
\newacronym{psd}{PSD}{power spectral density}
\newacronym{lvk}{LVK}{LIGO-Virgo-KAGRA Collaboration}
\newacronym{snr}{SNR}{signal-to-noise ratio}
\newacronym{gwtc}{GWTC}{Gravitational Wave Transient Catalog}
\newacronym{roq}{ROQ}{Reduced Order Quadrature}
\newacronym{ps}{PS}{Persistent Sampling}
\newacronym{pcn}{pCN}{preconditioned Crank-Nicolson}
\newacronym{tpcn}{tpCN}{$t$-preconditioned Crank-Nicolson}
\newacronym{ess}{ESS}{effective sample size}
\newacronym{pdf}{PDF}{probability density function}
\newacronym{kl}{KL divergence}{Kullback-Leibler divergence}
\newacronym{sbi}{SBI}{simulation-based inference}
\newacronym{rift}{RIFT}{Rapid Iterative FiTting}
\newacronym{aspire}{ASPIRE}{Accelerated Sequential Posterior Inference via Reuse}
\newacronym{maf}{MAF}{masked autoregressive flow}
\newacronym{kld}{KLD}{Kullback-Leibler divergence}

\newcommand{\bilby}{\texttt{bilby}\xspace}
\newcommand{\dynesty}{\texttt{dynesty}\xspace}
\newcommand{\minipcn}{\texttt{minipcn}\xspace}
\newcommand{\aspire}{\texttt{aspire}\xspace}
\newcommand{\jsd}{D_\textrm{JS}}

\newcommand{\ptemcee}{\texttt{ptemcee}\xspace}

\newcommand{\kld}{D_{\textrm{KL}}}
\newcommand{\hellinger}{D_{\textrm{H}^2}}

\begin{document}

\newcommand{\MaxEvalSpeedupSMC}{16.6\xspace}
\newcommand{\MaxTimeSpeedupSMC}{11.1\xspace}
\newcommand{\MaxEvalSpeedupTotal}{5.8\xspace}
\newcommand{\MaxTimeSpeedupTotal}{5.5\xspace}
\newcommand{\MaxEvalSpeedupPerSampleTotal}{12.3\xspace}
\newcommand{\MaxTimeSpeedupPerSampleTotal}{9.8\xspace}


\title{Accelerated Sequential Posterior Inference via Reuse for Gravitational-Wave Analyses}


\author{Michael J. Williams\,\orcidlink{0000-0003-2198-2974}}
\email[]{michael.williams3@port.ac.uk}
\affiliation{Institute of Cosmology and Gravitation, University of Portsmouth, Portsmouth PO1 3FX, United Kingdom}

\date{\today}

\begin{abstract}
We introduce \gls{aspire}, a broadly applicable framework that transforms existing posterior samples and Bayesian evidence estimates into unbiased results under alternative models without rerunning the original analysis. \Gls{aspire} combines normalizing flows with a generalized \gls{smc} scheme, enabling efficient updates of existing results and reducing total likelihood evaluations and wall times by factors of up to \MaxEvalSpeedupTotal and \MaxTimeSpeedupTotal, respectively, with larger gains per posterior sample. This addresses a growing problem in gravitational-wave astronomy, where events must be repeatedly reanalyzed under different models or physical hypotheses. We show that \gls{aspire} reproduces full Bayesian results when switching waveform models or adding physical effects such as spin precession and orbital eccentricity. With this statistical robustness, \gls{aspire} turns repeated reanalyses into fast, reliable updates---paving the way for systematic studies of waveform systematics, scalable reanalyses across large event catalogs, and broadly applicable Bayesian reanalysis across other scientific domains.
\end{abstract}

\glsresetall


\maketitle

\pdfbookmark[1]{Introduction}{sec:intro}
\noindent\emph{Introduction}---The LIGO--Virgo--KAGRA collaboration~\cite{LIGOScientific:2014pky,VIRGO:2014yos,KAGRA:2020tym} has detected more than 200 gravitational-wave signals from compact binary mergers~\cite{LIGOScientific:2018mvr,LIGOScientific:2020ibl,LIGOScientific:2021usb,KAGRA:2021vkt,LIGOScientific:2025slb} and continues to regularly detect such signals~\cite{LIGOG2302098}. These detections have shaped our understanding of the Universe on multiple scales~\cite{LIGOScientific:2021aug} and enabled tests of General Relativity~\cite{LIGOScientific:2025jau}. As the catalog of events grows, a key challenge emerges: events must be reanalyzed repeatedly under different models or physical hypotheses. Understanding waveform systematics~\cite{Gamba:2020wgg,Yelikar:2024wzm}, adding effects such as orbital eccentricity~\cite{Gamboa:2024hli,Nagar:2024dzj}, performing tests of General Relativity~\cite{Krishnendu:2021fga}, or modeling gravitational lensing~\cite{LIGOScientific:2021izm} all require new analyses from scratch. This repeated reanalysis is rapidly becoming a computational bottleneck, limiting our ability to probe waveform systematics and to conduct comprehensive tests of fundamental physics.

Ideally, existing results would inform subsequent analyses. Naively using a previous posterior as the prior for a new analysis double counts data which, if uncorrected, breaks model comparison via Bayes factors or posterior odds and leads to overly narrow posterior confidence intervals. Therefore, we require an algorithm that can incorporate existing samples while still producing unbiased posterior and Bayesian evidence estimates. Current analyses typically use nested sampling~\cite{Skilling:2004pqw,Skilling:2006gxv} which estimates both the posterior and evidence but can take hours to days~\cite{Veitch:2014wba,Romero-Shaw:2020owr}.
Many efforts have sought to directly accelerate inference~\cite{Williams:2021qyt,Williams:2023ppp,Chua:2019wwt,Gabbard:2019rde,Dax:2022pxd,Fairhurst:2023idl} but these methods do not leverage existing results. The standard approach for reusing posteriors within gravitational-wave inference is importance sampling~\cite{Ashton:2025xba}, however, this fails when the initial and target distributions differ substantially. More sophisticated methods include RIFT~\cite{Lange:2018pyp} and Posterior Repartitioning~\cite{Prathaban:2024rmu}, which can both be seeded from existing samples but require similar support to the posterior or fail when the distributions are disjoint~\cite{Prathaban:2026kft}.
Methods based on \gls{mcmc}~\cite{Hogg:2017akh} and \gls{ptmcmc}~\cite{2005PCCP....7.3910E} have also been explored~\cite{Ashton:2021anp,Wolfe:2022nkv,Falxa:2022yrm,Littenberg:2023xpl,Korsakova:2024sut}. While \gls{ptmcmc} can provide evidence estimates, gravitational-wave applications have not generally emphasized accurate evidence calculation, and robustness depends sensitively on choices such as the temperature ladder and mixing efficiency.

\begin{figure}
    \centering
    \includegraphics[width=\linewidth]{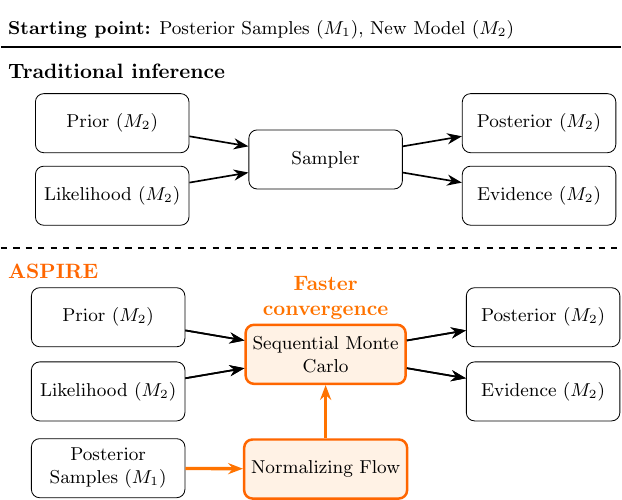}
    \caption{Comparison of traditional Bayesian inference (top), which must perform full inference for a new model $M_2$ despite existing posterior samples with $M_1$, with \gls{aspire} (bottom), which reuses the existing posterior samples under model $M_1$ via flow approximation and a Sequential Monte Carlo update to obtain unbiased results under a new model $M_2$.}
    \label{fig:algorithm}
\end{figure}

Here we present a novel framework: \gls{aspire}. Instead of discarding existing results and repeating full analyses, posterior samples and the evidence estimate obtained under one model are updated to unbiased posterior samples and an evidence estimate under a new model. This approach remains valid even when the two posteriors occupy different regions of parameter space or when new parameters are introduced, enabling extensions to additional physical effects (e.g., orbital eccentricity~\cite{Peters:1963ux}) and families of modified-gravity hypotheses~\cite{Krishnendu:2021fga}.
\Gls{aspire} thus enables repeated reanalyses via sequential reuse of existing results.

\Gls{aspire} combines normalizing flows with a generalized \gls{smc} scheme. We show that \gls{aspire} is robust and reproduces results that are statistically consistent with traditional analyses while reducing sampling time by up to \MaxTimeSpeedupTotal times. We demonstrate efficient updates between different waveform models, recovery of new parameters such as for modeling spin precession \cite{Apostolatos:1994mx} and orbital eccentricity \cite{Peters:1963ux}, and highlight an application to real data. By enabling analyses to be updated rather than repeated, \gls{aspire} provides a broadly applicable tool for gravitational-wave astronomy and other areas of physics where reanalyses are common.

\pdfbookmark[1]{Method}{sec:method}
\noindent\emph{Method}---Bayesian inference traditionally proceeds by assuming a model $M_1$ that describes the data $d$ and has parameters $\theta$. It then combines a prior $p(\theta|M_1)$ with a likelihood $p(d|\theta, M_1)$ to obtain a posterior distribution $p(\theta|M_1, d)$. Each time the model or data changes, the full inference is typically repeated from scratch (\cref{fig:algorithm}, top), which becomes costly as data volumes and model complexities grow.

The \gls{aspire} framework introduces a new approach to sequentially updating existing results and requires two complementary steps. Instead of restarting from draws from the prior, \gls{aspire} begins with posterior samples obtained under an existing model $M_1$ (\cref{fig:algorithm}, bottom left). We approximate these samples using a normalizing flow~\cite{Papamakarios:2019fms}, a flexible density estimator that maps a simple base distribution $q_z(z)$ into an approximation of the posterior  $q_{\phi}(\theta)$ through an invertible transformation $f_{\phi}(\theta)$ described by parameters $\phi$. Training the flow on the initial samples by maximum likelihood~\cite{Papamakarios:2019fms} yields
\begin{equation}
    p(\theta| d, M_1) \approx q_{\phi}(\theta) = q_z(f_{\phi}(\theta)) \left| \frac{\partial f_\phi(\theta)}{\partial \theta}\right|.
\end{equation}
This allows \gls{aspire} to start from an informed initial distribution rather than the prior. When $M_2$ has additional parameters, we extend the samples with draws from the prior and discard incompatible ones, ensuring that $q_\phi$ spans the parameter space of $M_2$ (see Supplemental Material~\cite{SM}).

The second step (\cref{fig:algorithm}, bottom middle) is to evolve this distribution to the target posterior and to estimate the Bayesian evidence. We build on \cite{Williams:2025szm} and generalize \gls{smc} for this purpose. \Gls{smc} for Bayesian inference evolves a set of samples through a sequence of intermediate distributions that interpolate between the prior and posterior via an inverse temperature $\beta_t$ that anneals the likelihood~\cite{moral2007sequential}. In \gls{aspire}, the sequence instead interpolates between the flow-based approximation $q_{\phi}(\theta)$ and the true posterior under the new model $M_2$:
\begin{equation}\label{eq:smc:p_t_aspire}
    p_t(\theta|d,M_2,\beta_t) \propto q_\phi(\theta)^{1-\beta_t}\,[p(d|\theta,M_2)p(\theta|M_2)]^{\beta_t},
\end{equation}
where $\beta_t \in [0, 1]$  controls the progression. The algorithm starts with samples from $q_{\phi}$ and then at each iteration $i$, the inverse temperature $\beta_i$ is increased using an adaptive schedule. The samples are resampled according to $p_i(\theta|d,M_2,\beta_i) / p_{i-1}(\theta|d,M_2,\beta_{i-1})$ and then diversified using an \gls{mcmc} step~\cite{del2012adaptive,Fearnhead:2013}. The algorithm terminates when $\beta_t=1$, producing posterior samples and an evidence estimate. This approach guarantees unbiased results, even when the target and initial posteriors occupy different regions of the parameter space, since the annealing schedule ensures that successive intermediate distributions have sufficient overlap and the resampling and mutation steps allow particles to move between regions (See fig.~4 of the Supplemental Material~\cite{SM} and the surrounding text).

Together, these two components, flow-based posterior approximation and generalized \gls{smc} evolution, define \gls{aspire}. This enables efficient reuse of existing results and allows analyses to be updated rather than repeated. The implementation details are provided in the Supplemental Material \cite{SM} and we provide an open-source implementation in the Python package \aspire~\cite{aspire-doi}.

\begin{figure}
    \centering
    \includegraphics[width=\linewidth]{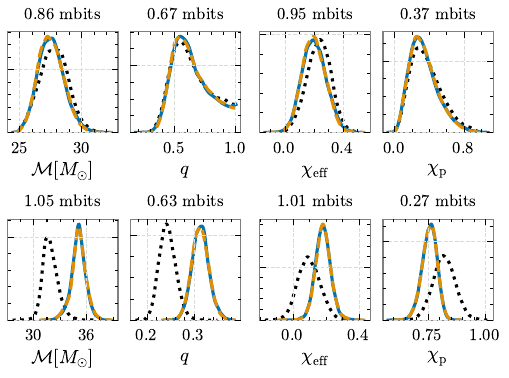}
    \caption{
    Posterior distributions for two simulated binary black hole signals analyzed with different waveform models. \textbf{Top:} GW150914-like system with comparable masses and low spins. \textbf{Bottom}: highly spinning asymmetric system with $m_1/m_2=4$. Results are shown for \texttt{IMRPhenomXPHM} with \dynesty (black dotted line),  \texttt{IMRPhenomXO4a} using \gls{aspire}, initialized from the \texttt{IMRPhenomXPHM} samples (orange dashed line) and \texttt{IMRPhenomXO4a} with \dynesty (blue solid line). Parameters shown are chirp mass, mass ratio, effective aligned spin $\chi_\mathrm{eff}$ and effective precessing spin $\chi_p$. The \gls{jsd} in millibits (mbits) between the two \texttt{IMRPhenomXO4a} posteriors is quoted above each subplot.
    }
    \label{fig:changing_waveform}
\end{figure}

\noindent\emph{Results}---We evaluate \gls{aspire} across simulated and real data, testing its robustness to waveform choice, added physical complexity, and detector calibration uncertainties.

We assess \gls{aspire} when changing the waveform model used to describe the signal, a key test to evaluate systematic uncertainties. We consider two waveform models that include spin precession and higher-order multipoles: \texttt{IMRPhenomXPHM}~\cite{Pratten:2020ceb} and \texttt{IMRPhenomXO4a}~\cite{Thompson:2023ase}, both used in GWTC-4.0~\cite{LIGOScientific:2025slb} to understand waveform systematics. \Cref{fig:changing_waveform} compares results for two simulated signals analyzed with both waveforms, where \gls{aspire} updates an initial \texttt{IMRPhenomXPHM} run to obtain results with \texttt{IMRPhenomXO4a}. We quantify the similarity between posterior distributions using the \gls{jsd} ($\jsd$) in millibits (mbits), where larger values indicate greater differences, $\jsd \le 50$ mbits denotes good agreement~\cite{LIGOScientific:2018mvr} and $\jsd \le 2.1$ mbits indicates the distributions are statistically indistinguishable~\cite{Romero-Shaw:2020owr}. For both systems, the posteriors obtained with \gls{aspire} are in close agreement with the baseline, even for the asymmetric binary where the posterior shifts substantially.

\begin{figure}
    \centering
    \includegraphics{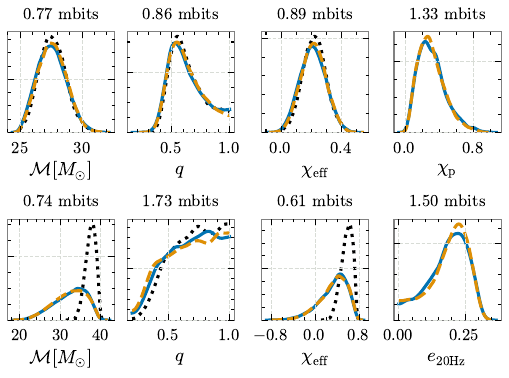}
    \caption{
        Posterior distributions when adding new physical effects to an initial analysis. \textbf{Top:} spin-precessing analysis of a GW150914-like injection, results are shown for the \texttt{IMRPhenomD} (aligned spins) baseline with \dynesty (black dotted line), \texttt{IMRPhenomPv2} (precessing spins) using \gls{aspire} initialized from the aligned-spin samples (orange dashed line) and an \texttt{IMRPhenomPv2} with \dynesty (blue solid line). \textbf{Bottom:} eccentric analysis of an aligned-spin GW150914-like injection, results are shown for the \texttt{TaylorF2} baseline with \dynesty (black dotted line), \texttt{TaylorF2Ecc} using \gls{aspire} initialized from the \texttt{TaylorF2} result (orange dashed line) and \texttt{TaylorF2Ecc} with \dynesty (blue solid line). Parameters shown are chirp mass, mass ratio, effective aligned spin $\chi_\mathrm{eff}$, and effective precessing spin $\chi_p$ (top) or eccentricity at 20 Hz $e_{20\textrm{Hz}}$ (bottom). The \gls{jsd} in millibits (mbits) between the posteriors with the same waveform is quoted above each subplot.}
    \label{fig:adding_eccentricity}
\end{figure}

We now consider analyses that incorporate additional physics, namely spin precession and orbital eccentricity. Starting with spin precession, we use \gls{aspire} to update a result obtained under the aligned-spin assumption with \texttt{IMRPhenomD}~\cite{Husa:2015iqa,Khan:2015jqa}, replacing it with the precessing model \texttt{IMRPhenomPv2}~\cite{Schmidt:2012rh,Hannam:2013oca}. This type of update is relevant in low-latency analyses, where simplifying assumptions are often made to reduce computational cost~\cite{Morisaki:2023kuq}. \Cref{fig:adding_eccentricity} (top) compares the posterior distributions and finds that \gls{aspire} is consistent with a baseline analysis with $\jsd < 1.5\;\textrm{mbits}$. This includes the precessing-spin parameter $\chi_p$, which captures physical effects absent from the initial aligned-spin analysis.

Orbital eccentricity is astrophysically interesting because it provides a clear signature of dynamical formation channels \cite{Mandel:2021smh} and several observed events show potential evidence for it \cite{Romero-Shaw:2020thy,Romero-Shaw:2022xko,Gupte:2024jfe,Morras:2025xfu}. However, incorporating eccentricity typically requires reanalyzing existing events, which can be computationally expensive.
To demonstrate robustness when adding this physics, we consider a simulated signal with $e_{20\,\mathrm{Hz}}=0.25$ and use \gls{aspire} to update a quasi-circular aligned-spin analysis with \texttt{TaylorF2}~\cite{Buonanno:2009zt} to the eccentric model \texttt{TaylorF2Ecc}~\cite{Moore:2016qxz}. \Cref{fig:adding_eccentricity} (bottom) shows the posterior distributions, including the eccentricity posterior, which agrees with the baseline analysis.

These examples illustrate that \gls{aspire} incorporates new physical degrees of freedom without introducing bias, even in cases where the added parameters are strongly correlated with existing ones.

\begin{figure}
    \centering
    \includegraphics[width=\linewidth]{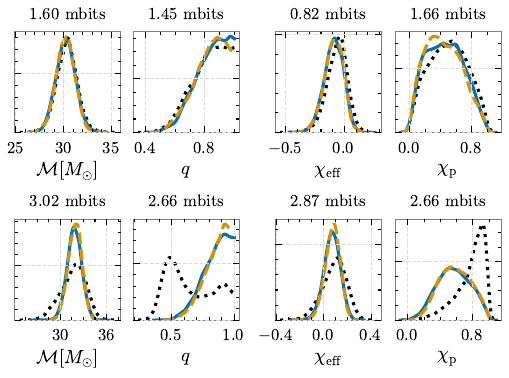}
    \caption{Posterior distributions for real gravitational-wave events analyzed with \gls{aspire}. \textbf{Top}: the first detection, GW150914. \textbf{Bottom}: an event with known waveform systematics, GW200129. Parameters shown are the chirp mass and mass ratio and aligned spin $\chi_\mathrm{eff}$ and effective precessing spin $\chi_\mathrm{p}$. Results are shown for an initial analysis with \texttt{IMRPhenomXPHM} and \dynesty (black dotted line), an analysis with \texttt{IMRPhenomXO4a} and \gls{aspire}, updated from the initial analysis (orange dashed line) and a baseline analysis with \texttt{IMRPhenomXO4a} and \dynesty (blue solid line).
    The \gls{jsd} in millibits (mbits) between the two \texttt{IMRPhenomXO4a} posteriors is quoted above each subplot.
    }
    \label{fig:real_data}
\end{figure}

We apply \gls{aspire} to real data by reanalyzing GW150914~\cite{LIGOScientific:2016vbw} and GW200129\_065458 from GWTC-3~\cite{KAGRA:2021vkt}, an event known to exhibit waveform systematics~\cite{KAGRA:2021vkt,Nitz:2021zwj,Hannam:2021pit,Payne:2022spz,Macas:2023wiw}.
We include calibration uncertainty~\cite{Viets:2017yvy} and follow a typical catalog workflow~\cite{LIGOScientific:2025yae}: first analyzing each event with \texttt{IMRPhenomXPHM}, then \texttt{IMRPhenomXO4a}.
GW150914 lies in a region of parameter space where these two waveform models are expected to agree broadly, whereas for GW200129 several parameters are known to be in tension between models.
\Cref{fig:real_data} shows the posterior distributions for the masses and spins of both events. For GW150914, the posteriors differ only slightly between waveform models, consistent with previous reanalyses~\cite{Pratten:2020fqn,Pompili:2023tna}. The \gls{aspire} results are consistent with a baseline analysis, as supported by the \gls{jsd}.
For GW200129, as expected, we observe substantial waveform-driven differences. Nonetheless, \gls{aspire} remains consistent with the baseline analysis, though with more stringent settings (Supplemental Material~\cite{SM}) and smaller efficiency gains (\cref{tab:evidences}).
Overall, these results demonstrate that \gls{aspire} remains effective on real data provided the initial analysis is sufficiently informative.

\begin{figure}
    \centering
    \includegraphics[width=\linewidth]{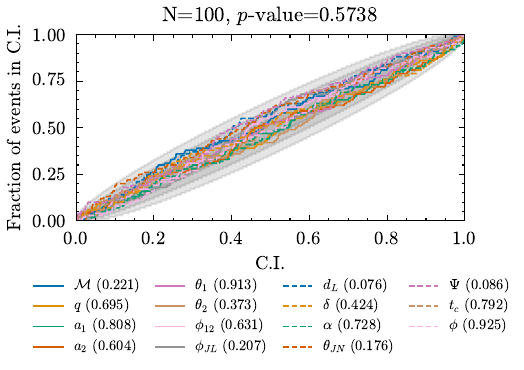}
    \caption{
    Probability–probability (P–P) plot for 100 simulated binary black hole signals analyzed with \gls{aspire}. Signals were generated with \texttt{IMRPhenomXPHM} in a three-detector network and analyzed using \bilby with \dynesty. Initial samples were drawn from the prior to initialize \gls{aspire}. Shaded regions indicate 1-, 2-, and 3-$\sigma$ confidence intervals. The combined $p$-value of 0.574 demonstrates that the posteriors are statistically unbiased.
    }
    \label{fig:pp}
\end{figure}

To further demonstrate that \gls{aspire} produces statistically unbiased posterior distributions, we perform a probability–probability (P–P) test with 100 simulated binary black hole signals with spin-precession and higher-order multipoles.  We analyze these signals with \gls{aspire} and use initial samples from the prior. This produces the P–P plot shown in \cref{fig:pp} with combined $p$-value of 0.574, confirming that the inferred posteriors are unbiased.

\begin{table*}
    \caption{Estimated log-evidences (base $e$), likelihood evaluations, wall time and number of posterior samples for analyses using \aspire and \dynesty. For \aspire, we separately report the initial SMC evolution cost and the total cost including posterior resampling. Analyses marked with $\dagger$ use more stringent settings (see Supplemental Material~\cite{SM}).}
    \label{tab:evidences}
    \begin{tabular}{p{2.8cm} c c c c c c c c}
\toprule
& \multicolumn{2}{c}{Log evidence} & \multicolumn{2}{c}{Likelihood evaluations [$10^6$]} & \multicolumn{2}{c}{Wall time [h]} & \multicolumn{2}{c}{Samples} \\
\cmidrule(lr){2-3} \cmidrule(lr){4-5} \cmidrule(lr){6-7} \cmidrule(lr){8-9}
Analysis & \texttt{dynesty} & \texttt{aspire} & \texttt{dynesty} & \texttt{aspire}& \texttt{dynesty} & \texttt{aspire} & \texttt{dynesty} & \texttt{aspire} \\
\midrule
\midrule
Changing waveform (GW150914-like) & $-57893.23 \pm 0.16$ & $-57893.36 \pm 0.06$ & 32.4 & 3.03 / 8.03 & 14.62 & 1.53 / 4.01 & 5810 & 10000 \\
Changing waveform ($q=4$)$^\dagger$ & $-54802.73 \pm 0.21$ & $-54803.15 \pm 0.07$ & 49.6 & 25.6 / 40.6 & 15.68 & 10.29 / 16.02 & 6511 & 20000 \\
Adding spin precession & $-57893.12 \pm 0.15$ & $-57893.11 \pm 0.06$ & 7.15 & 0.431 / 1.23 & 1.16 & 0.11 / 0.30 & 4737 & 10000 \\
Adding orbital eccentricity & $-1928.15 \pm 0.16$ & $-1928.26 \pm 0.06$ & 27.1 & 4.55 / 7.55 & 0.66 & 0.19 / 0.30 & 4240 & 10000 \\
GW150914 & $-12767.84 \pm 0.16$ & $-12768.08 \pm 0.07$ & 32.2 & 4.54 / 9.54 & 11.59 & 1.05 / 2.09 & 5627 & 10000 \\
GW200129$^\dagger$ & $-19556.51 \pm 0.19$ & $-19556.70 \pm 0.03$ & 53.6 & 37.7 / 52.7 & 19.45 & 11.24 / 13.93 & 7288 & 20000 \\
\bottomrule
\end{tabular}
\end{table*}

In addition to recovering unbiased posterior distributions, \gls{aspire} yields consistent evidence estimates. \Cref{tab:evidences} presents the inferred log-evidences for analyses using \gls{aspire} and \dynesty with the same waveform. The log-evidence differences between \gls{aspire} and \dynesty are within the estimated uncertainties, except for the asymmetric injection and GW200129 where they differ slightly. This agreement confirms that \gls{aspire} produces results consistent with baseline analyses and preserves statistical validity for Bayesian model comparison.

\Cref{tab:evidences} presents the number of likelihood evaluations, wall times, and posterior sample counts for the analyses. For \gls{aspire}, we separately report the cost of the initial \gls{smc} evolution and the total cost including final posterior resampling. The initial \gls{smc} stage governs the adaptation of the existing posterior to the target distribution, while additional posterior samples can be generated at low cost (see Supplemental Material~\cite{SM}). Across the scenarios considered here, \gls{aspire} reduces the cost of posterior adaptation by factors of up to \MaxEvalSpeedupSMC in likelihood evaluations and \MaxTimeSpeedupSMC in wall time. Including final posterior resampling, the total cost is reduced by up to \MaxEvalSpeedupTotal (\MaxEvalSpeedupPerSampleTotal per posterior sample), with corresponding wall time reductions of up to \MaxTimeSpeedupTotal (\MaxTimeSpeedupPerSampleTotal per posterior sample). More challenging updates, such as the asymmetric injection and GW200129, require more conservative settings, and therefore yield smaller gains.

\pdfbookmark[1]{Discussion and outlook}{sec:conclusions}
\noindent\emph{Discussion and Outlook}---\Gls{aspire} establishes a novel approach for Bayesian inference in gravitational-wave astronomy. Rather than discarding results and restarting analyses from the prior, posterior results can be transformed into unbiased results under new models, including new posterior samples and an evidence estimate that are consistent with traditional analyses.
By demonstrating that sequential inference via reuse of existing results is both practical and statistically robust, \gls{aspire} provides a new tool for systematic exploration of waveform systematics and physical hypotheses.

Our studies highlight the immediate utility of this approach. \Gls{aspire} reproduces full Bayesian results across waveform approximants, recovers additional physical effects such as spin precession and orbital eccentricity, and passes stringent statistical validation tests. In all cases, the method yields posteriors and evidence estimates consistent with baseline inferences while reducing the computational cost of repeated reanalyses, with reductions in total likelihood evaluations and wall times of up to \MaxEvalSpeedupTotal and \MaxTimeSpeedupTotal, respectively. When normalized by the number of posterior samples produced, the efficiency gains are even larger. These gains transform tasks that were previously prohibitive, such as repeated waveform comparisons or large-scale catalog updates, into routine analyses. In doing so, \gls{aspire} maximizes the scientific return from the rapidly growing catalog of gravitational-wave detections~\cite{LIGOScientific:2025slb}.
However, the efficiency of \gls{aspire} depends on the similarity between the initial and target posteriors. As this discrepancy increases, more \gls{smc} steps are required, reducing the computational advantage; in practice, known differences can often be mitigated by augmenting the initial samples with prior draws (Supplemental Material~\cite{SM}). Predicting performance a priori remains an open problem.

Several approaches have been proposed to accelerate or reuse existing inference results, including importance sampling~\cite{Ashton:2025xba}, nested-sampling initialisation strategies~\cite{Prathaban:2024rmu,Prathaban:2026kft}, and tempered \gls{mcmc} methods~\cite{2005PCCP....7.3910E,Vousden:2016ptemcee,Graham:2017,Wolfe:2022nkv,Falxa:2022yrm,Littenberg:2023xpl,Korsakova:2024sut}. 
Among these, \gls{ptmcmc} is the most directly comparable to \gls{aspire}. By evolving chains across a temperature ladder, \gls{ptmcmc} can explore multimodal structure and access regions not present in an initialization; however, its performance depends sensitively on temperature ladder design and mixing. 
In contrast, \gls{aspire} automatically constructs a sequence of intermediate distributions between a learned approximation and the target posterior, enabling targeted updates of existing results.
From the comparisons in the Supplemental Material~\cite{SM}, \gls{aspire} provides more efficient inference and more stable evidence estimates in the scenarios considered, while exhibiting limitations associated with its reliance on the support of the initial distribution.
These approaches therefore target different inference regimes.

Looking forward, the framework offers clear paths for further improvement. Future improvements, including gradient-based proposals~\cite{buchholz2021adaptive} and particle recycling~\cite{LeThuNguyen:2014,Karamanis:2025persistent}, could further accelerate \gls{aspire}. Since \gls{smc} is naturally parallelizable, the framework is well suited to GPU and low-latency applications.
Furthermore, alternative schemes for tempering and sequential updates could be explored~\cite{1998physics...3008N,Graham:2017,2024arXiv240812057S}. These developments position \gls{aspire} to benefit directly from advances in both algorithms and computing hardware.

Beyond methodological advances, \gls{aspire} has a wide range of potential applications. In gravitational-wave astronomy, it can support systematic waveform model comparisons, large-scale tests of General Relativity, rapid reanalyses of candidate events, and efficient updates of entire catalogs as models improve. It could be used to refine approximate posterior estimates from low-latency approaches~\cite{Dax:2022pxd,Fairhurst:2023idl,Nitz:2024nhj}. Early warning and low-latency pipelines~\cite{Morisaki:2023kuq,CabournDavies:2024hea} could incorporate \gls{aspire} to refine source characterization in real time. The framework also extends naturally to other fields where competing models abound, such as reinterpreting particle physics measurements under alternative theoretical models~\cite{Gartner:2024muk} or reanalysing cosmological data when performing Bayesian model comparison~\cite{Arjona:2021uxs,Lane:2023ndt}. By enabling fast and unbiased reanalyses, \gls{aspire} opens the way to scalable and systematic Bayesian inference across diverse scientific domains.

\begin{acknowledgments}

\noindent\emph{Acknowledgments}---The author thanks Colm Talbot and Konstantin Leyde for insightful discussions and Ian Harry and Chris Messenger for feedback on the manuscript. This work benefited from discussions at AIslands 2025: Bute (13--15 May 2025) and EuCAIFCon 2025 (16--20 June 2025). The author thanks the organizers and participants for their feedback and insights. The author acknowledges support from ST/X002225/1, ST/Y004876/1 and the University of Portsmouth. The author is grateful for computational resources provided by the LIGO Laboratory and Cardiff University and supported by National Science Foundation Grants PHY-0757058 and PHY-0823459 and STFC grants ST/I006285/1 and ST/V005618/1. Additional numerical computations were carried out on the SCIAMA High Performance Compute (HPC) cluster which is supported by the ICG and the University of Portsmouth. This material is based upon work supported by NSF's LIGO Laboratory which is a major facility fully funded by the National Science Foundation. The author acknowledges the use of ChatGPT (GPT-4-turbo and GPT-5) for proofreading and polishing the manuscript; all AI-generated text was carefully reviewed and edited by the author to ensure accuracy.
\end{acknowledgments}

\noindent\emph{Data availability}---The results presented in this letter and code to reproduce the analyses are available at \cite{data_release,code_release}. The \aspire~\cite{aspire-doi}, \texttt{aspire-bilby}~\cite{aspire-bilby-doi} and \texttt{aspire-gw}~\cite{aspire-gw-doi} software packages are available via PyPI~\cite{pypi-aspire,pypi-aspire-bilby,pypi-aspire-gw} and the source code at \url{https://github.com/mj-will/aspire}, \url{https://github.com/mj-will/aspire-bilby} and \url{https://github.com/mj-will/aspire-gw}.
Interferometric data for GW150914 and GW200129 is provided by the LIGO-Virgo-KAGRA Collaboration via GWOSC \cite{LIGOScientific:2019lzm}.

\noindent\emph{Software}---This work used the following software packages:
\texttt{array-api-compat}~\cite{ArrayAPICompat}, 
\bilby~\cite{Ashton:2018jfp, bilby}, 
\texttt{bilby\_pipe}~\cite{Romero-Shaw:2020owr}, 
\texttt{corner}~\cite{corner}, 
\texttt{dynesty}~\cite{Speagle:2019ivv}, 
\texttt{gwflow}~\cite{gwflow-doi},
\texttt{LALSuite}~\cite{lalsuite,swiglal}, 
\texttt{matplotlib}~\cite{Hunter:2007}, 
\minipcn~\cite{minipcn}, 
\texttt{numpy}~\cite{numpy}, 
\texttt{PESummary}~\cite{Hoy:2020vys},
\texttt{ptemcee}~\cite{emcee,Vousden:2016ptemcee},
\texttt{PyTorch}~\cite{Paszke:2019xhz}, 
\texttt{scipy}~\cite{mckinney-proc-scipy-2010,2020SciPy-NMeth}, 
\texttt{seaborn}~\cite{Waskom2021}, 
\texttt{zuko}~\cite{zuko}.

\nocite{Karamanis:2022ksp,Vaart_1998,douc2005comparison,2024InvPr..40l5023G,Bose:2021pcw,Papamakarios:2017tec,flowjax,Kingma:2014vow,Loshchilov:2016ofz,KAGRA:2013rdx,Baird:2012cu,Thrane:2018qnx,Kullback:1951zyt,Hellinger1909,ANNIS201967,Gelman:1992zz,nessai}

\bibliography{ref.bib}

@PREAMBLE{
 "\providecommand{\noopsort}[1]{}" 
 # "\providecommand{\singleletter}[1]{#1}%" 
}

@misc{SM,
  note = {See Supplemental Material for additional implementation and analysis details.},
}

@misc{Williams:2025szm,
    author = "Williams, Michael J. and Karamanis, Minas and Luo, Yilin and Seljak, Uro{\v{s}}",
    title = "{Validating Sequential Monte Carlo for Gravitational-Wave Inference}",
    eprint = "2506.18977",
    archivePrefix = "arXiv",
    primaryClass = "astro-ph.IM",
    reportNumber = "LIGO-P2500231",
    month = "6",
    year = "2025"
}

@misc{Papamakarios:2017tec,
    author = "Papamakarios, George and Pavlakou, Theo and Murray, Iain",
    title = "{Masked Autoregressive Flow for Density Estimation}",
    eprint = "1705.07057",
    archivePrefix = "arXiv",
    primaryClass = "stat.ML",
    month = "5",
    year = "2017"
}

@article{LIGOScientific:2018mvr,
    author = "Abbott, B. P. and others",
    collaboration = "LIGO Scientific, Virgo",
    title = "{GWTC-1: A Gravitational-Wave Transient Catalog of Compact Binary Mergers Observed by LIGO and Virgo during the First and Second Observing Runs}",
    eprint = "1811.12907",
    archivePrefix = "arXiv",
    primaryClass = "astro-ph.HE",
    reportNumber = "LIGO-P1800307",
    doi = "10.1103/PhysRevX.9.031040",
    journal = "Phys. Rev. X",
    volume = "9",
    number = "3",
    pages = "031040",
    year = "2019"
}

@article{LIGOScientific:2020ibl,
    author = "Abbott, R. and others",
    collaboration = "LIGO Scientific, Virgo",
    title = "{GWTC-2: Compact Binary Coalescences Observed by LIGO and Virgo During the First Half of the Third Observing Run}",
    eprint = "2010.14527",
    archivePrefix = "arXiv",
    primaryClass = "gr-qc",
    reportNumber = "P2000061",
    doi = "10.1103/PhysRevX.11.021053",
    journal = "Phys. Rev. X",
    volume = "11",
    pages = "021053",
    year = "2021"
}

@article{LIGOScientific:2021usb,
    author = "Abbott, R. and others",
    collaboration = "LIGO Scientific, VIRGO",
    title = "{GWTC-2.1: Deep extended catalog of compact binary coalescences observed by LIGO and Virgo during the first half of the third observing run}",
    eprint = "2108.01045",
    archivePrefix = "arXiv",
    primaryClass = "gr-qc",
    reportNumber = "LIGO-P2100063",
    doi = "10.1103/PhysRevD.109.022001",
    journal = "Phys. Rev. D",
    volume = "109",
    number = "2",
    pages = "022001",
    year = "2024"
}

@article{KAGRA:2021vkt,
    author = "Abbott, R. and others",
    collaboration = "KAGRA, VIRGO, LIGO Scientific",
    title = "{GWTC-3: Compact Binary Coalescences Observed by LIGO and Virgo during the Second Part of the Third Observing Run}",
    eprint = "2111.03606",
    archivePrefix = "arXiv",
    primaryClass = "gr-qc",
    reportNumber = "LIGO-P2000318",
    doi = "10.1103/PhysRevX.13.041039",
    journal = "Phys. Rev. X",
    volume = "13",
    number = "4",
    pages = "041039",
    year = "2023"
}

@article{LIGOScientific:2014pky,
    author = "Aasi, J. and others",
    collaboration = "LIGO Scientific",
    title = "{Advanced LIGO}",
    eprint = "1411.4547",
    archivePrefix = "arXiv",
    primaryClass = "gr-qc",
    doi = "10.1088/0264-9381/32/7/074001",
    journal = "Class. Quant. Grav.",
    volume = "32",
    pages = "074001",
    year = "2015"
}

@article{VIRGO:2014yos,
    author = "Acernese, F. and others",
    collaboration = "VIRGO",
    title = "{Advanced Virgo: a second-generation interferometric gravitational wave detector}",
    eprint = "1408.3978",
    archivePrefix = "arXiv",
    primaryClass = "gr-qc",
    doi = "10.1088/0264-9381/32/2/024001",
    journal = "Class. Quant. Grav.",
    volume = "32",
    number = "2",
    pages = "024001",
    year = "2015"
}

@article{KAGRA:2020tym,
    author = "Akutsu, T. and others",
    collaboration = "KAGRA",
    title = "{Overview of KAGRA: Detector design and construction history}",
    eprint = "2005.05574",
    archivePrefix = "arXiv",
    primaryClass = "physics.ins-det",
    doi = "10.1093/ptep/ptaa125",
    journal = "PTEP",
    volume = "2021",
    number = "5",
    pages = "05A101",
    year = "2021"
}

@misc{LIGOG2302098,
    author = {{LIGO Scientific Collaboration}},
    title = {{LIGO-G2302098: LIGO-Virgo-KAGRA Cumulative Detection plot - O1-O4b}},
    howpublished = {\url{https://dcc.ligo.org/LIGO-G2302098-v15/public}},
    year = {2025},
    note = {Accessed: July 7, 2025}
}

@article{LIGOScientific:2021aug,
    author = "Abbott, R. and others",
    collaboration = "LIGO Scientific, Virgo, KAGRA",
    title = "{Constraints on the Cosmic Expansion History from GWTC{\textendash}3}",
    eprint = "2111.03604",
    archivePrefix = "arXiv",
    primaryClass = "astro-ph.CO",
    reportNumber = "LIGO-P2100185-v6, LIGO-P2100185-v5",
    doi = "10.3847/1538-4357/ac74bb",
    journal = "Astrophys. J.",
    volume = "949",
    number = "2",
    pages = "76",
    year = "2023"
}

@article{Skilling:2004pqw,
    author = "Skilling, John",
    title = "{Nested Sampling}",
    doi = "10.1063/1.1835238",
    journal = "AIP Conf. Proc.",
    volume = "735",
    number = "1",
    pages = "395",
    year = "2004"
}

@article{Skilling:2006gxv,
    author = "Skilling, John",
    title = "{Nested sampling for general Bayesian computation}",
    doi = "10.1214/06-BA127",
    journal = "Bayesian Analysis",
    volume = "1",
    number = "4",
    pages = "833--859",
    year = "2006"
}

@article{KAGRA:2013rdx,
    author = "Abbott, B. P. and others",
    collaboration = "KAGRA, LIGO Scientific, Virgo",
    title = "{Prospects for observing and localizing gravitational-wave transients with Advanced LIGO, Advanced Virgo and KAGRA}",
    eprint = "1304.0670",
    archivePrefix = "arXiv",
    primaryClass = "gr-qc",
    reportNumber = "LIGO-P1200087, VIR-0288A-12, JGW-P1808427",
    doi = "10.1007/s41114-020-00026-9",
    journal = "Living Rev. Rel.",
    volume = "19",
    pages = "1",
    year = "2016"
}

@misc{zuko,
  title = {{Zuko}: Normalizing flows in PyTorch},
  author = {Rozet, François and others},
  year = {2024},
  doi = {10.5281/zenodo.7625672},
  license = {MIT},
  url = {https://pypi.org/project/zuko},
}

@misc{flowjax,
  title = {FlowJAX: Distributions and Normalizing Flows in Jax},
  author = {Daniel Ward},
  url = {https://github.com/danielward27/flowjax},
  year = {2025},
  doi = {10.5281/zenodo.10402073},
}

@article{Moore:2016qxz,
    author = "Moore, Blake and Favata, Marc and Arun, K. G. and Mishra, Chandra Kant",
    title = "{Gravitational-wave phasing for low-eccentricity inspiralling compact binaries to 3PN order}",
    eprint = "1605.00304",
    archivePrefix = "arXiv",
    primaryClass = "gr-qc",
    reportNumber = "LIGO-DCC-P1500268",
    doi = "10.1103/PhysRevD.93.124061",
    journal = "Phys. Rev. D",
    volume = "93",
    number = "12",
    pages = "124061",
    year = "2016"
}

@article{Buonanno:2009zt,
    author = "Buonanno, Alessandra and Iyer, Bala and Ochsner, Evan and Pan, Yi and Sathyaprakash, B. S.",
    title = "{Comparison of post-Newtonian templates for compact binary inspiral signals in gravitational-wave detectors}",
    eprint = "0907.0700",
    archivePrefix = "arXiv",
    primaryClass = "gr-qc",
    doi = "10.1103/PhysRevD.80.084043",
    journal = "Phys. Rev. D",
    volume = "80",
    pages = "084043",
    year = "2009"
}

@article{Gamba:2020wgg,
    author = "Gamba, Rossella and Breschi, Matteo and Bernuzzi, Sebastiano and Agathos, Michalis and Nagar, Alessandro",
    title = "{Waveform systematics in the gravitational-wave inference of tidal parameters and equation of state from binary neutron star signals}",
    eprint = "2009.08467",
    archivePrefix = "arXiv",
    primaryClass = "gr-qc",
    doi = "10.1103/PhysRevD.103.124015",
    journal = "Phys. Rev. D",
    volume = "103",
    number = "12",
    pages = "124015",
    year = "2021"
}

@article{Yelikar:2024wzm,
    author = "Yelikar, A. B. and Shaughnessy, R. O' and Lange, J. and Jan, A. Z.",
    title = "{Waveform systematics in gravitational-wave inference of signals from binary neutron star merger models incorporating higher-order modes information}",
    eprint = "2404.16599",
    archivePrefix = "arXiv",
    primaryClass = "gr-qc",
    reportNumber = "LIGO DCC P2400138",
    doi = "10.1103/PhysRevD.110.064024",
    journal = "Phys. Rev. D",
    volume = "110",
    number = "6",
    pages = "064024",
    year = "2024"
}

@article{LIGOScientific:2021izm,
    author = "Abbott, R. and others",
    collaboration = "LIGO Scientific, VIRGO",
    title = "{Search for Lensing Signatures in the Gravitational-Wave Observations from the First Half of LIGO{\textendash}Virgo{\textquoteright}s Third Observing Run}",
    eprint = "2105.06384",
    archivePrefix = "arXiv",
    primaryClass = "gr-qc",
    reportNumber = "LIGO-P2000400",
    doi = "10.3847/1538-4357/ac23db",
    journal = "Astrophys. J.",
    volume = "923",
    number = "1",
    pages = "14",
    year = "2021"
}

@article{Krishnendu:2021fga,
    author = "Krishnendu, N. V. and Ohme, Frank",
    title = "{Testing General Relativity with Gravitational Waves: An Overview}",
    eprint = "2201.05418",
    archivePrefix = "arXiv",
    primaryClass = "gr-qc",
    doi = "10.3390/universe7120497",
    journal = "Universe",
    volume = "7",
    number = "12",
    pages = "497",
    year = "2021"
}

@article{Williams:2021qyt,
    author = "Williams, Michael J. and Veitch, John and Messenger, Chris",
    title = "{Nested sampling with normalizing flows for gravitational-wave inference}",
    eprint = "2102.11056",
    archivePrefix = "arXiv",
    primaryClass = "gr-qc",
    doi = "10.1103/PhysRevD.103.103006",
    journal = "Phys. Rev. D",
    volume = "103",
    number = "10",
    pages = "103006",
    year = "2021"
}

@article{Williams:2023ppp,
    author = "Williams, Michael J. and Veitch, John and Messenger, Chris",
    title = "{Importance nested sampling with normalising flows}",
    eprint = "2302.08526",
    archivePrefix = "arXiv",
    primaryClass = "astro-ph.IM",
    reportNumber = "LIGO-P2200283",
    doi = "10.1088/2632-2153/acd5aa",
    journal = "Mach. Learn. Sci. Tech.",
    volume = "4",
    number = "3",
    pages = "035011",
    year = "2023"
}

@article{Dax:2022pxd,
    author = {Dax, Maximilian and Green, Stephen R. and Gair, Jonathan and P{\"u}rrer, Michael and Wildberger, Jonas and Macke, Jakob H. and Buonanno, Alessandra and Sch{\"o}lkopf, Bernhard},
    title = "{Neural Importance Sampling for Rapid and Reliable Gravitational-Wave Inference}",
    eprint = "2210.05686",
    archivePrefix = "arXiv",
    primaryClass = "gr-qc",
    reportNumber = "LIGO-P2200297",
    doi = "10.1103/PhysRevLett.130.171403",
    journal = "Phys. Rev. Lett.",
    volume = "130",
    number = "17",
    pages = "171403",
    year = "2023"
}

@article{Fairhurst:2023idl,
    author = "Fairhurst, Stephen and Hoy, Charlie and Green, Rhys and Mills, Cameron and Usman, Samantha A.",
    title = "{Simple parameter estimation using observable features of gravitational-wave signals}",
    eprint = "2304.03731",
    archivePrefix = "arXiv",
    primaryClass = "gr-qc",
    doi = "10.1103/PhysRevD.108.082006",
    journal = "Phys. Rev. D",
    volume = "108",
    number = "8",
    pages = "082006",
    year = "2023"
}

@misc{Prathaban:2024rmu,
    author = "Prathaban, Metha and Bevins, Harry and Handley, Will",
    title = "{Accelerated nested sampling with $\beta$-flows for gravitational waves}",
    eprint = "2411.17663",
    archivePrefix = "arXiv",
    primaryClass = "astro-ph.IM",
    month = "11",
    year = "2024"
}

@article{Papamakarios:2019fms,
    author = "Papamakarios, George and Nalisnick, Eric and Rezende, Danilo Jimenez and Mohamed, Shakir and Lakshminarayanan, Balaji",
    title = "{Normalizing Flows for Probabilistic Modeling and Inference}",
    eprint = "1912.02762",
    archivePrefix = "arXiv",
    primaryClass = "stat.ML",
    doi = "10.5555/3546258.3546315",
    journal = "J. Machine Learning Res.",
    volume = "22",
    number = "1",
    pages = "2617--2680",
    year = "2021"
}

@incollection{moral2007sequential,
    author = {Moral, Pierre Del and Doucet, Arnaud and Jasra, Ajay},
    editor = {Bernardo, J M and Bayarri, M J and Berger, J O and Dawid, A P and Heckerman, D and Smith, A F M and West, M and Bernardo, J M and Bayarri, M J and Berger, J O and Dawid, A P and Heckerman, D and Smith, A F M and West, M},
    isbn = {9780199214655},
    title = {Sequential Monte Carlo for Bayesian Computation},
    booktitle = {Bayesian Statistics 8: Proceedings of the Eighth Valencia International Meeting June 2–6, 2006},
    publisher = {Oxford University Press},
    year = {2007},
    month = {07},
    doi = {10.1093/oso/9780199214655.003.0005},
    url = {https://doi.org/10.1093/oso/9780199214655.003.0005},
    eprint = {https://academic.oup.com/book/0/chapter/422208917/chapter-pdf/52446914/isbn-9780199214655-book-part-5.pdf},
}

@article{del2012adaptive,
  title={An adaptive sequential Monte Carlo method for approximate Bayesian computation},
  author={Del Moral, Pierre and Doucet, Arnaud and Jasra, Ajay},
  journal={Statistics and computing},
  volume={22},
  pages={1009--1020},
  year={2012},
  publisher={Springer}
}

@article{Fearnhead:2013,
author = {Paul Fearnhead and Benjamin M. Taylor},
title = {{An Adaptive Sequential Monte Carlo Sampler}},
volume = {8},
journal = {Bayesian Analysis},
number = {2},
publisher = {International Society for Bayesian Analysis},
pages = {411 -- 438},
year = {2013},
doi = {10.1214/13-BA814},
URL = {https://doi.org/10.1214/13-BA814}
}

@article{Thompson:2023ase,
    author = "Thompson, Jonathan E. and Hamilton, Eleanor and London, Lionel and Ghosh, Shrobana and Kolitsidou, Panagiota and Hoy, Charlie and Hannam, Mark",
    title = "{PhenomXO4a: a phenomenological gravitational-wave model for precessing black-hole binaries with higher multipoles and asymmetries}",
    eprint = "2312.10025",
    archivePrefix = "arXiv",
    primaryClass = "gr-qc",
    reportNumber = "LIGO-P2300437",
    doi = "10.1103/PhysRevD.109.063012",
    journal = "Phys. Rev. D",
    volume = "109",
    number = "6",
    pages = "063012",
    year = "2024"
}

@article{Pratten:2020ceb,
    author = "Pratten, Geraint and others",
    title = "{Computationally efficient models for the dominant and subdominant harmonic modes of precessing binary black holes}",
    eprint = "2004.06503",
    archivePrefix = "arXiv",
    primaryClass = "gr-qc",
    doi = "10.1103/PhysRevD.103.104056",
    journal = "Phys. Rev. D",
    volume = "103",
    number = "10",
    pages = "104056",
    year = "2021"
}

@article{Karamanis:2022ksp,
    author = "Karamanis, Minas and Nabergoj, David and Beutler, Florian and Peacock, John A. and Seljak, Uros",
    title = "{pocoMC: A Python package for accelerated Bayesian inference in astronomy and cosmology}",
    eprint = "2207.05660",
    archivePrefix = "arXiv",
    primaryClass = "astro-ph.IM",
    doi = "10.21105/joss.04634",
    journal = "J. Open Source Softw.",
    volume = "7",
    number = "79",
    pages = "4634",
    year = "2022"
}

@ARTICLE{2024InvPr..40l5023G,
       author = {{Grumitt}, Richard D.~P. and {Karamanis}, Minas and {Seljak}, Uro{\v{s}}},
        title = "{Sequential Kalman tuning of the t-preconditioned Crank-Nicolson algorithm: efficient, adaptive and gradient-free inference for Bayesian inverse problems}",
      journal = {Inverse Problems},
         year = 2024,
        month = dec,
       volume = {40},
       number = {12},
          eid = {125023},
        pages = {125023},
          doi = {10.1088/1361-6420/ad934b},
archivePrefix = {arXiv},
       eprint = {2407.07781},
 primaryClass = {stat.CO},
       adsurl = {https://ui.adsabs.harvard.edu/abs/2024InvPr..40l5023G}
}

@misc{minipcn,
  author       = {Michael J. Williams},
  title        = {mj-will/minipcn},
  month        = jun,
  year         = 2025,
  publisher    = {Zenodo},
  version      = {v0.1.0b1},
  doi          = {10.5281/zenodo.15657998},
  url          = {https://doi.org/10.5281/zenodo.15657998},
}

@inproceedings{Kingma:2014vow,
    author = "Kingma, Diederik P. and Ba, Jimmy",
    title = "{Adam: A Method for Stochastic Optimization}",
    eprint = "1412.6980",
    archivePrefix = "arXiv",
    primaryClass = "cs.LG",
    month = "12",
    year = "2014"
}

@misc{Loshchilov:2016ofz,
    author = "Loshchilov, Ilya and Hutter, Frank",
    title = "{SGDR: Stochastic Gradient Descent with Warm Restarts}",
    eprint = "1608.03983",
    archivePrefix = "arXiv",
    primaryClass = "cs.LG",
    year = "2016",
    month = "Aug",
    doi = "10.48550/arXiv.1608.03983"
}

@article{Thrane:2018qnx,
    author = "Thrane, Eric and Talbot, Colm",
    title = "{An introduction to Bayesian inference in gravitational-wave astronomy: parameter estimation, model selection, and hierarchical models}",
    eprint = "1809.02293",
    archivePrefix = "arXiv",
    primaryClass = "astro-ph.IM",
    doi = "10.1017/pasa.2019.2",
    journal = "Publ. Astron. Soc. Austral.",
    volume = "36",
    pages = "e010",
    year = "2019",
    note = "[Erratum: Publ.Astron.Soc.Austral. 37, e036 (2020)]"
}

@misc{Gamboa:2024hli,
    author = "Gamboa, Aldo and others",
    title = "{Accurate waveforms for eccentric, aligned-spin binary black holes: The multipolar effective-one-body model SEOBNRv5EHM}",
    eprint = "2412.12823",
    archivePrefix = "arXiv",
    primaryClass = "gr-qc",
    month = "12",
    year = "2024"
}

@article{Nagar:2024dzj,
    author = "Nagar, Alessandro and Gamba, Rossella and Rettegno, Piero and Fantini, Veronica and Bernuzzi, Sebastiano",
    title = "{Effective-one-body waveform model for noncircularized, planar, coalescing black hole binaries: The importance of radiation reaction}",
    eprint = "2404.05288",
    archivePrefix = "arXiv",
    primaryClass = "gr-qc",
    doi = "10.1103/PhysRevD.110.084001",
    journal = "Phys. Rev. D",
    volume = "110",
    number = "8",
    pages = "084001",
    year = "2024"
}

@article{Hogg:2017akh,
    author = "Hogg, David W. and Foreman-Mackey, Daniel",
    title = "{Data analysis recipes: Using Markov Chain Monte Carlo}",
    eprint = "1710.06068",
    archivePrefix = "arXiv",
    primaryClass = "astro-ph.IM",
    doi = "10.3847/1538-4365/aab76e",
    journal = "Astrophys. J. Suppl.",
    volume = "236",
    number = "1",
    pages = "11",
    year = "2018"
}

@article{Husa:2015iqa,
    author = {Husa, Sascha and Khan, Sebastian and Hannam, Mark and P{\"u}rrer, Michael and Ohme, Frank and Jim{\'e}nez Forteza, Xisco and Boh{\'e}, Alejandro},
    title = "{Frequency-domain gravitational waves from nonprecessing black-hole binaries. I. New numerical waveforms and anatomy of the signal}",
    eprint = "1508.07250",
    archivePrefix = "arXiv",
    primaryClass = "gr-qc",
    doi = "10.1103/PhysRevD.93.044006",
    journal = "Phys. Rev. D",
    volume = "93",
    number = "4",
    pages = "044006",
    year = "2016"
}

@article{Khan:2015jqa,
    author = {Khan, Sebastian and Husa, Sascha and Hannam, Mark and Ohme, Frank and P{\"u}rrer, Michael and Jim{\'e}nez Forteza, Xisco and Boh{\'e}, Alejandro},
    title = "{Frequency-domain gravitational waves from nonprecessing black-hole binaries. II. A phenomenological model for the advanced detector era}",
    eprint = "1508.07253",
    archivePrefix = "arXiv",
    primaryClass = "gr-qc",
    doi = "10.1103/PhysRevD.93.044007",
    journal = "Phys. Rev. D",
    volume = "93",
    number = "4",
    pages = "044007",
    year = "2016"
}

@article{Schmidt:2012rh,
    author = "Schmidt, Patricia and Hannam, Mark and Husa, Sascha",
    title = "{Towards models of gravitational waveforms from generic binaries: A simple approximate mapping between precessing and non-precessing inspiral signals}",
    eprint = "1207.3088",
    archivePrefix = "arXiv",
    primaryClass = "gr-qc",
    doi = "10.1103/PhysRevD.86.104063",
    journal = "Phys. Rev. D",
    volume = "86",
    pages = "104063",
    year = "2012"
}

@article{Hannam:2013oca,
    author = {Hannam, Mark and Schmidt, Patricia and Boh{\'e}, Alejandro and Haegel, Le{\"\i}la and Husa, Sascha and Ohme, Frank and Pratten, Geraint and P{\"u}rrer, Michael},
    title = "{Simple Model of Complete Precessing Black-Hole-Binary Gravitational Waveforms}",
    eprint = "1308.3271",
    archivePrefix = "arXiv",
    primaryClass = "gr-qc",
    doi = "10.1103/PhysRevLett.113.151101",
    journal = "Phys. Rev. Lett.",
    volume = "113",
    number = "15",
    pages = "151101",
    year = "2014"
}

@misc{bilby,
  author       = {Colm Talbot and
                  Gregory Ashton and
                  Moritz Hübner and
                  Matt Pitkin and
                  plasky and
                  asb5468 and
                  Michael J. Williams and
                  Rory Smith and
                  Aditya Vijaykumar and
                  SMorisaki and
                  John Veitch and
                  Nikhil Sarin and
                  Duncan Macleod and
                  Daniel Williams and
                  MarcArene and
                  JasperMartins and
                  C P L Berry and
                  Vivien Raymond and
                  Ceciliogq and
                  Ivan Markin and
                  David Keitel and
                  AlexandreGoettel and
                  Lorenzo Pompili and
                  Mick Wright and
                  oliviawilk and
                  noahewolfe and
                  jacobgolomb and
                  Shichao Wu and
                  Rhiannon Udall and
                  Michael Pürrer},
  title        = {bilby-dev/bilby: v2.5.2},
  month        = jun,
  year         = 2025,
  publisher    = {Zenodo},
  version      = {v2.5.2},
  doi          = {10.5281/zenodo.15586392},
  url          = {https://doi.org/10.5281/zenodo.15586392},
  swhid        = {swh:1:dir:51e86257c7c1d8ab1b3bf3597a31dab5c62fdd06
                   ;origin=https://doi.org/10.5281/zenodo.14025463;vi
                   sit=swh:1:snp:7d942bcad0d82cdbaac4343d7e4a71a64f3e
                   336d;anchor=swh:1:rel:556a1cb0e50e6ea0e6a381ffeee3
                   03c0c224b7b8;path=bilby-dev-bilby-33b5cf0
                  },
}

@article{Ashton:2018jfp,
    author = "Ashton, Gregory and others",
    title = "{BILBY: A user-friendly Bayesian inference library for gravitational-wave astronomy}",
    eprint = "1811.02042",
    archivePrefix = "arXiv",
    primaryClass = "astro-ph.IM",
    doi = "10.3847/1538-4365/ab06fc",
    journal = "Astrophys. J. Suppl.",
    volume = "241",
    number = "2",
    pages = "27",
    year = "2019"
}

@article{Speagle:2019ivv,
    author = "Speagle, Joshua S.",
    title = "{dynesty: a dynamic nested sampling package for estimating Bayesian posteriors and evidences}",
    eprint = "1904.02180",
    archivePrefix = "arXiv",
    primaryClass = "astro-ph.IM",
    doi = "10.1093/mnras/staa278",
    journal = "Mon. Not. Roy. Astron. Soc.",
    volume = "493",
    number = "3",
    pages = "3132--3158",
    year = "2020"
}

@article{Romero-Shaw:2020owr,
    author = "Romero-Shaw, I. M. and others",
    title = "{Bayesian inference for compact binary coalescences with bilby: validation and application to the first LIGO{\textendash}Virgo gravitational-wave transient catalogue}",
    eprint = "2006.00714",
    archivePrefix = "arXiv",
    primaryClass = "astro-ph.IM",
    doi = "10.1093/mnras/staa2850",
    journal = "Mon. Not. Roy. Astron. Soc.",
    volume = "499",
    number = "3",
    pages = "3295--3319",
    year = "2020"
}

@ARTICLE{2020SciPy-NMeth,
  author  = {Virtanen, Pauli and Gommers, Ralf and Oliphant, Travis E. and
            Haberland, Matt and Reddy, Tyler and Cournapeau, David and
            Burovski, Evgeni and Peterson, Pearu and Weckesser, Warren and
            Bright, Jonathan and {van der Walt}, St{\'e}fan J. and
            Brett, Matthew and Wilson, Joshua and Millman, K. Jarrod and
            Mayorov, Nikolay and Nelson, Andrew R. J. and Jones, Eric and
            Kern, Robert and Larson, Eric and Carey, C J and
            Polat, {\.I}lhan and Feng, Yu and Moore, Eric W. and
            {VanderPlas}, Jake and Laxalde, Denis and Perktold, Josef and
            Cimrman, Robert and Henriksen, Ian and Quintero, E. A. and
            Harris, Charles R. and Archibald, Anne M. and
            Ribeiro, Ant{\^o}nio H. and Pedregosa, Fabian and
            {van Mulbregt}, Paul and {SciPy 1.0 Contributors}},
  title   = {{{SciPy} 1.0: Fundamental Algorithms for Scientific
            Computing in Python}},
  journal = {Nature Methods},
  year    = {2020},
  volume  = {17},
  pages   = {261--272},
  adsurl  = {https://rdcu.be/b08Wh},
  doi     = {10.1038/s41592-019-0686-2},
}

@misc{lalsuite,
       author         = "{LIGO Scientific Collaboration}",
       title          = "{LIGO} {A}lgorithm {L}ibrary - {LALS}uite",
       howpublished   = "free software (GPL)",
       doi            = "10.7935/GT1W-FZ16",
       year           = "2018"
 }

@article{swiglal,
          title     = "{SWIGLAL: Python and Octave interfaces to the LALSuite gravitational-wave data analysis libraries}",
          author    = "Karl Wette",
          journal   = "SoftwareX",
          volume    = "12",
          pages     = "100634",
          year      = "2020",
          doi       = "10.1016/j.softx.2020.100634"
 }

@Article{Hunter:2007,
  Author    = {Hunter, J. D.},
  Title     = {Matplotlib: A 2D graphics environment},
  Journal   = {Computing in Science \& Engineering},
  Volume    = {9},
  Number    = {3},
  Pages     = {90--95},
  publisher = {IEEE COMPUTER SOC},
  doi       = {10.1109/MCSE.2007.55},
  year      = 2007
}

@InProceedings{ mckinney-proc-scipy-2010,
  author    = { {W}es {M}c{K}inney },
  title     = { {D}ata {S}tructures for {S}tatistical {C}omputing in {P}ython },
  booktitle = { {P}roceedings of the 9th {P}ython in {S}cience {C}onference },
  pages     = { 56 - 61 },
  year      = { 2010 },
  editor    = { {S}t\'efan van der {W}alt and {J}arrod {M}illman },
  doi       = { 10.25080/Majora-92bf1922-00a }
}

@Article{numpy,
 title         = {Array programming with {NumPy}},
 author        = {Charles R. Harris and K. Jarrod Millman and St{\'{e}}fan J.
                 van der Walt and Ralf Gommers and Pauli Virtanen and David
                 Cournapeau and Eric Wieser and Julian Taylor and Sebastian
                 Berg and Nathaniel J. Smith and Robert Kern and Matti Picus
                 and Stephan Hoyer and Marten H. van Kerkwijk and Matthew
                 Brett and Allan Haldane and Jaime Fern{\'{a}}ndez del
                 R{\'{i}}o and Mark Wiebe and Pearu Peterson and Pierre
                 G{\'{e}}rard-Marchant and Kevin Sheppard and Tyler Reddy and
                 Warren Weckesser and Hameer Abbasi and Christoph Gohlke and
                 Travis E. Oliphant},
 year          = {2020},
 month         = sep,
 journal       = {Nature},
 volume        = {585},
 number        = {7825},
 pages         = {357--362},
 doi           = {10.1038/s41586-020-2649-2},
 publisher     = {Springer Science and Business Media {LLC}},
 url           = {https://doi.org/10.1038/s41586-020-2649-2}
}

@article{corner,
      doi = {10.21105/joss.00024},
      url = {https://doi.org/10.21105/joss.00024},
      year  = {2016},
      month = {6},
      publisher = {The Open Journal},
      volume = {1},
      number = {2},
      pages = {24},
      author = {Daniel Foreman-Mackey},
      title = {corner.py: Scatterplot matrices in Python},
      journal = {The Journal of Open Source Software}
    }

@article{Hoy:2020vys,
    author = "Hoy, Charlie and Raymond, Vivien",
    title = "{PESummary: the code agnostic Parameter Estimation Summary page builder}",
    eprint = "2006.06639",
    archivePrefix = "arXiv",
    primaryClass = "astro-ph.IM",
    reportNumber = "LIGO-P2000156",
    doi = "10.1016/j.softx.2021.100765",
    journal = "SoftwareX",
    volume = "15",
    pages = "100765",
    year = "2021"
}

@article{Waskom2021,
    doi = {10.21105/joss.03021},
    url = {https://doi.org/10.21105/joss.03021},
    year = {2021},
    publisher = {The Open Journal},
    volume = {6},
    number = {60},
    pages = {3021},
    author = {Michael L. Waskom},
    title = {seaborn: statistical data visualization},
    journal = {Journal of Open Source Software}
 }

@misc{Paszke:2019xhz,
    author = "Paszke, Adam and others",
    title = "{PyTorch: An Imperative Style, High-Performance Deep Learning Library}",
    eprint = "1912.01703",
    archivePrefix = "arXiv",
    primaryClass = "cs.LG",
    month = "12",
    year = "2019"
}

@misc{ArrayAPICompat,
    author = {{Consortium for Python Data API Standards}},
    title = "array-api-compat",
    year = 2025,
    howpublished = {\url{https://pypi.org/project/array-api-compat/}},
}

@misc{Lange:2018pyp,
    author = "Lange, Jacob and O'Shaughnessy, Richard and Rizzo, Monica",
    title = "{Rapid and accurate parameter inference for coalescing, precessing compact binaries}",
    eprint = "1805.10457",
    archivePrefix = "arXiv",
    primaryClass = "gr-qc",
    reportNumber = "LIGO DCC P1800084, LIGO-DCC-P1800084",
    month = "5",
    year = "2018"
}

@misc{LIGOScientific:2025slb,
    collaboration = "LIGO Scientific, VIRGO, KAGRA",
    title = "{GWTC-4.0: Updating the Gravitational-Wave Transient Catalog with Observations from the First Part of the Fourth LIGO-Virgo-KAGRA Observing Run}",
    eprint = "2508.18082",
    archivePrefix = "arXiv",
    primaryClass = "gr-qc",
    reportNumber = "LIGO-P2400386",
    month = "8",
    year = "2025"
}

@article{Veitch:2014wba,
    author = "Veitch, J. and others",
    title = "{Parameter estimation for compact binaries with ground-based gravitational-wave observations using the LALInference software library}",
    eprint = "1409.7215",
    archivePrefix = "arXiv",
    primaryClass = "gr-qc",
    reportNumber = "LIGO-P1400152",
    doi = "10.1103/PhysRevD.91.042003",
    journal = "Phys. Rev. D",
    volume = "91",
    number = "4",
    pages = "042003",
    year = "2015"
}

@article{Gabbard:2019rde,
    author = "Gabbard, Hunter and Messenger, Chris and Heng, Ik Siong and Tonolini, Francesco and Murray-Smith, Roderick",
    title = "{Bayesian parameter estimation using conditional variational autoencoders for gravitational-wave astronomy}",
    eprint = "1909.06296",
    archivePrefix = "arXiv",
    primaryClass = "astro-ph.IM",
    doi = "10.1038/s41567-021-01425-7",
    journal = "Nature Phys.",
    volume = "18",
    number = "1",
    pages = "112--117",
    year = "2022"
}

@article{buchholz2021adaptive,
  title={Adaptive tuning of hamiltonian monte carlo within sequential monte carlo},
  author={Buchholz, Alexander and Chopin, Nicolas and Jacob, Pierre E},
  journal={Bayesian Analysis},
  volume={16},
  number={3},
  pages={745--771},
  year={2021},
  publisher={International Society for Bayesian Analysis}
}

@INPROCEEDINGS{LeThuNguyen:2014,
  author={Le Thu Nguyen, Thi and Septier, Francois and Peters, Gareth W. and Delignon, Yves},
  booktitle={2014 IEEE Workshop on Statistical Signal Processing (SSP)}, 
  title={Improving SMC sampler estimate by recycling all past simulated particles}, 
  year={2014},
  volume={},
  number={},
  pages={117-120},
  doi={10.1109/SSP.2014.6884589}
}

@article{Karamanis:2025persistent,
  title={Persistent Sampling: Enhancing the Efficiency of Sequential Monte Carlo},
  author={Karamanis, Minas and Seljak, Uro{\v{s}}},
  journal={Statistics and Computing},
  volume={35},
  number={5},
  pages={1--22},
  year={2025},
  publisher={Springer}
}

@misc{pypi-aspire,
  title        = {aspire-inference},
  author       = {Williams, Michael J.},
  year         = {2025},
  howpublished = {Python Package Index (PyPI)},
  url          = {https://pypi.org/project/aspire-inference/},
  urldate      = {2025-09-22}
}

@misc{pypi-aspire-bilby,
  title        = {aspire-bilby},
  author       = {Williams, Michael J.},
  year         = {2025},
  howpublished = {Python Package Index (PyPI)},
  url          = {https://pypi.org/project/aspire-bilby/},
  urldate      = {2025-09-22}
}

@misc{pypi-aspire-gw,
  title        = {aspire-gw},
  author       = {Williams, Michael J.},
  year         = {2025},
  howpublished = {Python Package Index (PyPI)},
  url          = {https://pypi.org/project/aspire-gw/},
  urldate      = {2025-09-22}
}

@article{CabournDavies:2024hea,
    author = "Cabourn Davies, Gareth and others",
    title = "{Premerger observation and characterization of massive black hole binaries}",
    eprint = "2411.07020",
    archivePrefix = "arXiv",
    primaryClass = "hep-ex",
    doi = "10.1103/PhysRevD.111.043045",
    journal = "Phys. Rev. D",
    volume = "111",
    number = "4",
    pages = "043045",
    year = "2025"
}

@article{Morisaki:2023kuq,
    author = "Morisaki, Soichiro and Smith, Rory and Tsukada, Leo and Sachdev, Surabhi and Stevenson, Simon and Talbot, Colm and Zimmerman, Aaron",
    title = "{Rapid localization and inference on compact binary coalescences with the Advanced LIGO-Virgo-KAGRA gravitational-wave detector network}",
    eprint = "2307.13380",
    archivePrefix = "arXiv",
    primaryClass = "gr-qc",
    doi = "10.1103/PhysRevD.108.123040",
    journal = "Phys. Rev. D",
    volume = "108",
    number = "12",
    pages = "123040",
    year = "2023"
}

@article{Pratten:2020fqn,
    author = "Pratten, Geraint and Husa, Sascha and Garcia-Quiros, Cecilio and Colleoni, Marta and Ramos-Buades, Antoni and Estelles, Hector and Jaume, Rafel",
    title = "{Setting the cornerstone for a family of models for gravitational waves from compact binaries: The dominant harmonic for nonprecessing quasicircular black holes}",
    eprint = "2001.11412",
    archivePrefix = "arXiv",
    primaryClass = "gr-qc",
    reportNumber = "LIGO-P2000018",
    doi = "10.1103/PhysRevD.102.064001",
    journal = "Phys. Rev. D",
    volume = "102",
    number = "6",
    pages = "064001",
    year = "2020"
}

@article{Pompili:2023tna,
    author = "Pompili, Lorenzo and others",
    title = "{Laying the foundation of the effective-one-body waveform models SEOBNRv5: Improved accuracy and efficiency for spinning nonprecessing binary black holes}",
    eprint = "2303.18039",
    archivePrefix = "arXiv",
    primaryClass = "gr-qc",
    doi = "10.1103/PhysRevD.108.124035",
    journal = "Phys. Rev. D",
    volume = "108",
    number = "12",
    pages = "124035",
    year = "2023"
}

@misc{LIGOScientific:2025yae,
    author = "Abac, A. G. and others",
    collaboration = "LIGO Scientific, VIRGO, KAGRA",
    title = "{GWTC-4.0: Methods for Identifying and Characterizing Gravitational-wave Transients}",
    eprint = "2508.18081",
    archivePrefix = "arXiv",
    primaryClass = "gr-qc",
    reportNumber = "LIGO-P2400300",
    month = "8",
    year = "2025"
}

@article{Romero-Shaw:2022xko,
    author = "Romero-Shaw, Isobel M. and Lasky, Paul D. and Thrane, Eric",
    title = "{Four Eccentric Mergers Increase the Evidence that LIGO{\textendash}Virgo{\textendash}KAGRA{\textquoteright}s Binary Black Holes Form Dynamically}",
    eprint = "2206.14695",
    archivePrefix = "arXiv",
    primaryClass = "astro-ph.HE",
    doi = "10.3847/1538-4357/ac9798",
    journal = "Astrophys. J.",
    volume = "940",
    number = "2",
    pages = "171",
    year = "2022"
}

@article{Chua:2019wwt,
    author = "Chua, Alvin J. K. and Vallisneri, Michele",
    title = "{Learning Bayesian posteriors with neural networks for gravitational-wave inference}",
    eprint = "1909.05966",
    archivePrefix = "arXiv",
    primaryClass = "gr-qc",
    doi = "10.1103/PhysRevLett.124.041102",
    journal = "Phys. Rev. Lett.",
    volume = "124",
    number = "4",
    pages = "041102",
    year = "2020"
}

@misc{Gupte:2024jfe,
    author = "Gupte, Nihar and others",
    title = "{Evidence for eccentricity in the population of binary black holes observed by LIGO-Virgo-KAGRA}",
    eprint = "2404.14286",
    archivePrefix = "arXiv",
    primaryClass = "gr-qc",
    month = "4",
    year = "2024"
}

@misc{Morras:2025xfu,
    author = "Morras, Gonzalo and Pratten, Geraint and Schmidt, Patricia",
    title = "{Orbital eccentricity in a neutron star - black hole binary}",
    eprint = "2503.15393",
    archivePrefix = "arXiv",
    primaryClass = "astro-ph.HE",
    reportNumber = "LIGO-DCC P2500105",
    month = "3",
    year = "2025"
}

@article{Romero-Shaw:2020thy,
    author = "Romero-Shaw, Isobel M. and Lasky, Paul D. and Thrane, Eric and Bustillo, Juan Calderon",
    title = "{GW190521: orbital eccentricity and signatures of dynamical formation in a binary black hole merger signal}",
    eprint = "2009.04771",
    archivePrefix = "arXiv",
    primaryClass = "astro-ph.HE",
    doi = "10.3847/2041-8213/abbe26",
    journal = "Astrophys. J. Lett.",
    volume = "903",
    number = "1",
    pages = "L5",
    year = "2020"
}

@article{Viets:2017yvy,
    author = "Viets, Aaron and others",
    title = "{Reconstructing the calibrated strain signal in the Advanced LIGO detectors}",
    eprint = "1710.09973",
    archivePrefix = "arXiv",
    primaryClass = "astro-ph.IM",
    doi = "10.1088/1361-6382/aab658",
    journal = "Class. Quant. Grav.",
    volume = "35",
    number = "9",
    pages = "095015",
    year = "2018"
}

@article{Mandel:2021smh,
    author = "Mandel, Ilya and Broekgaarden, Floor S.",
    title = "{Rates of compact object coalescences}",
    eprint = "2107.14239",
    archivePrefix = "arXiv",
    primaryClass = "astro-ph.HE",
    doi = "10.1007/s41114-021-00034-3",
    journal = "Living Rev. Rel.",
    volume = "25",
    number = "1",
    pages = "1",
    year = "2022"
}

@misc{aspire-doi,
  author       = {Michael J. Williams and
                  Angel Garron and
                  Filippo Santoliquido and
                  Jacopo Tissino and
                  Konstantin Leyde},
  title        = {mj-will/aspire: v0.1.0a19},
  month        = mar,
  year         = 2026,
  publisher    = {Zenodo},
  version      = {v0.1.0a19},
  doi          = {10.5281/zenodo.19224148},
  url          = {https://doi.org/10.5281/zenodo.19224148},
  swhid        = {swh:1:dir:0e1c978089d86171bf37ef5102d5cba3fb2beddd
                   ;origin=https://doi.org/10.5281/zenodo.15658747;vi
                   sit=swh:1:snp:69520906d7825d7425a9e74fe709a1280648
                   3018;anchor=swh:1:rel:705af73d863307cf568cb1d9f80a
                   3af1cf7f662f;path=mj-will-aspire-2debbdd
                  },
}

@misc{gwflow-doi,
  author       = {Michael J. Williams},
  title        = {mj-will/gwflow: v0.1.0a1},
  month        = sep,
  year         = 2025,
  publisher    = {Zenodo},
  version      = {v0.1.0a1},
  doi          = {10.5281/zenodo.17182521},
  url          = {https://doi.org/10.5281/zenodo.17182521},
}

@misc{aspire-gw-doi,
  author       = {Michael J. Williams},
  title        = {mj-will/aspire-gw: v0.1.0a1},
  month        = sep,
  year         = 2025,
  publisher    = {Zenodo},
  version      = {v0.1.0a1},
  doi          = {10.5281/zenodo.17183944},
  url          = {https://doi.org/10.5281/zenodo.17183944},
}

@misc{aspire-bilby-doi,
  author       = {Michael J. Williams},
  title        = {mj-will/aspire-bilby: v0.1.0a2},
  month        = nov,
  year         = 2025,
  publisher    = {Zenodo},
  version      = {v0.1.0a2},
  doi          = {10.5281/zenodo.17540602},
  url          = {https://doi.org/10.5281/zenodo.17540602},
}

@misc{LIGOScientific:2025jau,
    collaboration = "LIGO Scientific, VIRGO, KAGRA",
    title = "{GWTC-4.0: Constraints on the Cosmic Expansion Rate and Modified Gravitational-wave Propagation}",
    eprint = "2509.04348",
    archivePrefix = "arXiv",
    primaryClass = "astro-ph.CO",
    reportNumber = "LIGO-P2400152",
    month = "9",
    year = "2025"
}

@article{LIGOScientific:2019lzm,
    author = "Abbott, Rich and others",
    collaboration = "LIGO Scientific, Virgo",
    title = "{Open data from the first and second observing runs of Advanced LIGO and Advanced Virgo}",
    eprint = "1912.11716",
    archivePrefix = "arXiv",
    primaryClass = "gr-qc",
    reportNumber = "LIGO-P1900206",
    doi = "10.1016/j.softx.2021.100658",
    journal = "SoftwareX",
    volume = "13",
    pages = "100658",
    year = "2021"
}

@article{LIGOScientific:2016vbw,
    author = "Abbott, B. P. and others",
    collaboration = "LIGO Scientific, Virgo",
    title = "{GW150914: First results from the search for binary black hole coalescence with Advanced LIGO}",
    eprint = "1602.03839",
    archivePrefix = "arXiv",
    primaryClass = "gr-qc",
    reportNumber = "LIGO-P1500269",
    doi = "10.1103/PhysRevD.93.122003",
    journal = "Phys. Rev. D",
    volume = "93",
    number = "12",
    pages = "122003",
    year = "2016"
}

@article{Apostolatos:1994mx,
    author = "Apostolatos, Theocharis A. and Cutler, Curt and Sussman, Gerald J. and Thorne, Kip S.",
    title = "{Spin induced orbital precession and its modulation of the gravitational wave forms from merging binaries}",
    reportNumber = "GRP-382",
    doi = "10.1103/PhysRevD.49.6274",
    journal = "Phys. Rev. D",
    volume = "49",
    pages = "6274--6297",
    year = "1994"
}

@article{Peters:1963ux,
    author = "Peters, P. C. and Mathews, J.",
    title = "{Gravitational radiation from point masses in a Keplerian orbit}",
    doi = "10.1103/PhysRev.131.435",
    journal = "Phys. Rev.",
    volume = "131",
    pages = "435--439",
    year = "1963"
}

@article{Wolfe:2022nkv,
    author = "Wolfe, Noah E. and Talbot, Colm and Golomb, Jacob",
    title = "{Accelerating tests of general relativity with gravitational-wave signals using hybrid sampling}",
    eprint = "2208.12872",
    archivePrefix = "arXiv",
    primaryClass = "gr-qc",
    doi = "10.1103/PhysRevD.107.104056",
    journal = "Phys. Rev. D",
    volume = "107",
    number = "10",
    pages = "104056",
    year = "2023"
}

@ARTICLE{Graham:2017,
       author = {{Graham}, Matthew M. and {Storkey}, Amos J.},
        title = "{Continuously tempered Hamiltonian Monte Carlo}",
      journal = {arXiv e-prints},
         year = 2017,
        month = apr,
          eid = {arXiv:1704.03338},
        pages = {arXiv:1704.03338},
          doi = {10.48550/arXiv.1704.03338},
archivePrefix = {arXiv},
       eprint = {1704.03338},
 primaryClass = {stat.CO},
       adsurl = {https://ui.adsabs.harvard.edu/abs/2017arXiv170403338G}
}

@article{Gartner:2024muk,
    author = {G{\"a}rtner, Lorenz and Hartmann, Nikolai and Heinrich, Lukas and Horstmann, Malin and Kuhr, Thomas and Reboud, M{\'e}ril and Stefkova, Slavomira and van Dyk, Danny},
    title = "{Constructing model-agnostic likelihoods, a method for the reinterpretation of particle physics results}",
    eprint = "2402.08417",
    archivePrefix = "arXiv",
    primaryClass = "hep-ph",
    reportNumber = "IPPP/24/06",
    doi = "10.1140/epjc/s10052-024-13038-4",
    journal = "Eur. Phys. J. C",
    volume = "84",
    number = "7",
    pages = "693",
    year = "2024"
}

@article{Arjona:2021uxs,
    author = "Arjona, Rub{\'e}n and Espinosa-Portales, Llorenc and Garc{\'\i}a-Bellido, Juan and Nesseris, Savvas",
    title = "{A GREAT model comparison against the cosmological constant}",
    eprint = "2111.13083",
    archivePrefix = "arXiv",
    primaryClass = "astro-ph.CO",
    reportNumber = "IFT-UAM/CSIC-2021-136",
    doi = "10.1016/j.dark.2022.101029",
    journal = "Phys. Dark Univ.",
    volume = "36",
    pages = "101029",
    year = "2022"
}

@article{Lane:2023ndt,
    author = "Lane, Zachary G. and Seifert, Antonia and Ridden-Harper, Ryan and Wiltshire, David L.",
    title = "{Cosmological foundations revisited with Pantheon+}",
    eprint = "2311.01438",
    archivePrefix = "arXiv",
    primaryClass = "astro-ph.CO",
    doi = "10.1093/mnras/stae2437",
    journal = "Mon. Not. Roy. Astron. Soc.",
    volume = "536",
    number = "2",
    pages = "1752--1777",
    year = "2025"
}

@misc{data_release,
  author       = {Williams, Michael J.},
  title        = "{Accelerated Sequential Posterior Inference via
                   Reuse for Gravitational-Wave Analyses - Data
                   Release
                  }",
  month        = nov,
  year         = 2025,
  publisher    = {Zenodo},
  version      = {v1.0},
  doi          = {10.5281/zenodo.17514969},
  url          = {https://doi.org/10.5281/zenodo.17514969},
}

@misc{code_release,
  author       = {Williams, Michael J.},
  title        = "{Accelerated Sequential Posterior Inference via
                   Reuse for Gravitational-Wave Analyses - Code
                   Release
                  }",
  month        = nov,
  year         = 2025,
  publisher    = {Zenodo},
  version      = {v1.0-alpha.2},
  doi          = {10.5281/zenodo.17536499},
  url          = {https://doi.org/10.5281/zenodo.17536499},
}

@article{Nitz:2024nhj,
    author = "Nitz, Alexander Harvey",
    title = "{Robust, rapid, and simple gravitational-wave parameter estimation}",
    eprint = "2410.05190",
    archivePrefix = "arXiv",
    primaryClass = "astro-ph.IM",
    doi = "10.1103/rml9-qyw1",
    journal = "Phys. Rev. D",
    volume = "112",
    number = "2",
    pages = "023032",
    year = "2025"
}

@article{Macas:2023wiw,
    author = "Macas, Ronaldas and Lundgren, Andrew and Ashton, Gregory",
    title = "{Revisiting the evidence for precession in GW200129 with machine learning noise mitigation}",
    eprint = "2311.09921",
    archivePrefix = "arXiv",
    primaryClass = "gr-qc",
    doi = "10.1103/PhysRevD.109.062006",
    journal = "Phys. Rev. D",
    volume = "109",
    number = "6",
    pages = "062006",
    year = "2024"
}

@article{Payne:2022spz,
    author = "Payne, Ethan and Hourihane, Sophie and Golomb, Jacob and Udall, Rhiannon and Udall, Richard and Davis, Derek and Chatziioannou, Katerina",
    title = "{Curious case of GW200129: Interplay between spin-precession inference and data-quality issues}",
    eprint = "2206.11932",
    archivePrefix = "arXiv",
    primaryClass = "gr-qc",
    reportNumber = "LIGO DCC: P2200185",
    doi = "10.1103/PhysRevD.106.104017",
    journal = "Phys. Rev. D",
    volume = "106",
    number = "10",
    pages = "104017",
    year = "2022"
}

@article{Nitz:2021zwj,
    author = {Nitz, Alexander H. and Kumar, Sumit and Wang, Yi-Fan and Kastha, Shilpa and Wu, Shichao and Sch{\"a}fer, Marlin and Dhurkunde, Rahul and Capano, Collin D.},
    title = "{4-OGC: Catalog of Gravitational Waves from Compact Binary Mergers}",
    eprint = "2112.06878",
    archivePrefix = "arXiv",
    primaryClass = "astro-ph.HE",
    doi = "10.3847/1538-4357/aca591",
    journal = "Astrophys. J.",
    volume = "946",
    number = "2",
    pages = "59",
    year = "2023"
}

@article{Hannam:2021pit,
    author = "Hannam, Mark and others",
    title = "{General-relativistic precession in a black-hole binary}",
    eprint = "2112.11300",
    archivePrefix = "arXiv",
    primaryClass = "gr-qc",
    reportNumber = "LIGO-P2100452",
    doi = "10.1038/s41586-022-05212-z",
    journal = "Nature",
    volume = "610",
    number = "7933",
    pages = "652--655",
    year = "2022"
}

@article{Ashton:2021anp,
    author = "Ashton, Gregory and Talbot, Colm",
    title = "{B{\,}ilby-MCMC: an MCMC sampler for gravitational-wave inference}",
    eprint = "2106.08730",
    archivePrefix = "arXiv",
    primaryClass = "gr-qc",
    doi = "10.1093/mnras/stab2236",
    journal = "Mon. Not. Roy. Astron. Soc.",
    volume = "507",
    number = "2",
    pages = "2037--2051",
    year = "2021"
}

@ARTICLE{emcee,
       author = {{Foreman-Mackey}, Daniel and {Hogg}, David W. and {Lang}, Dustin and {Goodman}, Jonathan},
        title = "{emcee: The MCMC Hammer}",
      journal = {Publications of the Astronomical Society of the Pacific},
         year = 2013,
        month = mar,
       volume = {125},
       number = {925},
        pages = {306},
          doi = {10.1086/670067},
archivePrefix = {arXiv},
       eprint = {1202.3665},
 primaryClass = {astro-ph.IM},
       adsurl = {https://ui.adsabs.harvard.edu/abs/2013PASP..125..306F}
}

@ARTICLE{Vousden:2016ptemcee,
       author = {{Vousden}, W.~D. and {Farr}, W.~M. and {Mandel}, I.},
        title = "{Dynamic temperature selection for parallel tempering in Markov chain Monte Carlo simulations}",
      journal = {Monthly Notices of the Royal Astronomical Society},
         year = 2016,
        month = jan,
       volume = {455},
       number = {2},
        pages = {1919-1937},
          doi = {10.1093/mnras/stv2422},
archivePrefix = {arXiv},
       eprint = {1501.05823},
 primaryClass = {astro-ph.IM},
       adsurl = {https://ui.adsabs.harvard.edu/abs/2016MNRAS.455.1919V}
}

@article{ANNIS201967,
title = {Thermodynamic integration and steppingstone sampling methods for estimating Bayes factors: A tutorial},
journal = {Journal of Mathematical Psychology},
volume = {89},
pages = {67-86},
year = {2019},
issn = {0022-2496},
doi = {https://doi.org/10.1016/j.jmp.2019.01.005},
url = {https://www.sciencedirect.com/science/article/pii/S0022249617302651},
author = {Jeffrey Annis and Nathan J. Evans and Brent J. Miller and Thomas J. Palmeri}
}

@inproceedings{douc2005comparison,
  title={Comparison of resampling schemes for particle filtering},
  author={Douc, Randal and Capp{\'e}, Olivier},
  booktitle={ISPA 2005. Proceedings of the 4th International Symposium on Image and Signal Processing and Analysis, 2005.},
  pages={64--69},
  year={2005},
  organization={Ieee}
}

@ARTICLE{2005PCCP....7.3910E,
       author = {{Earl}, David J. and {Deem}, Michael W.},
        title = "{Parallel tempering: Theory, applications, and new perspectives}",
      journal = {Physical Chemistry Chemical Physics (Incorporating Faraday Transactions)},
         year = 2005,
        month = jan,
       volume = {7},
       number = {23},
        pages = {3910},
          doi = {10.1039/B509983H},
archivePrefix = {arXiv},
       eprint = {physics/0508111},
 primaryClass = {physics.comp-ph},
       adsurl = {https://ui.adsabs.harvard.edu/abs/2005PCCP....7.3910E}
}

@inbook{Vaart_1998, place={Cambridge}, series={Cambridge Series in Statistical and Probabilistic Mathematics}, title={Delta Method}, booktitle={Asymptotic Statistics}, publisher={Cambridge University Press}, author={Vaart, A. W. van der}, year={1998}, pages={25–34}, collection={Cambridge Series in Statistical and Probabilistic Mathematics}}

@misc{Ashton:2025xba,
    author = "Ashton, Gregory",
    title = "{Reconstructing and resampling: a guide to utilising posterior samples from gravitational wave observations}",
    eprint = "2510.11197",
    archivePrefix = "arXiv",
    primaryClass = "gr-qc",
    month = "10",
    year = "2025"
}

@misc{Prathaban:2026kft,
    author = "Prathaban, Metha and Hoy, Charlie and Williams, Michael J.",
    title = "{Leveraging rapid parameter estimates for efficient gravitational-wave Bayesian inference via posterior repartitioning}",
    eprint = "2601.21630",
    archivePrefix = "arXiv",
    primaryClass = "gr-qc",
    reportNumber = "LIGO-P2600020",
    month = "1",
    year = "2026"
}

@article{Hellinger1909,
author = {Hellinger, E.},
journal = {Journal für die reine und angewandte Mathematik},
pages = {210-271},
title = {Neue Begründung der Theorie quadratischer Formen von unendlichvielen Veränderlichen.},
url = {http://eudml.org/doc/149313},
volume = {136},
year = {1909},
}

@article{Kullback:1951zyt,
    author = "Kullback, S. and Leibler, R. A.",
    title = "{On Information and Sufficiency}",
    doi = "10.1214/aoms/1177729694",
    journal = "The Annals of Mathematical Statistics",
    volume = "22",
    number = "1",
    pages = "79--86",
    year = "1951"
}

@article{Korsakova:2024sut,
    author = "Korsakova, Natalia and Babak, Stanislav and Katz, Michael L. and Karnesis, Nikolaos and Khukhlaev, Sviatoslav and Gair, Jonathan R.",
    title = "{Neural density estimation for Galactic binaries in the LISA data analysis}",
    eprint = "2402.13701",
    archivePrefix = "arXiv",
    primaryClass = "gr-qc",
    doi = "10.1103/PhysRevD.110.104069",
    journal = "Phys. Rev. D",
    volume = "110",
    number = "10",
    pages = "104069",
    year = "2024"
}

@article{Falxa:2022yrm,
    author = "Falxa, Mikel and Babak, Stanislav and Le Jeune, Maude",
    title = "{Adaptive kernel density estimation proposal in gravitational wave data analysis}",
    eprint = "2208.04575",
    archivePrefix = "arXiv",
    primaryClass = "astro-ph.IM",
    doi = "10.1103/PhysRevD.107.022008",
    journal = "Phys. Rev. D",
    volume = "107",
    number = "2",
    pages = "022008",
    year = "2023"
}

@article{Littenberg:2023xpl,
    author = "Littenberg, Tyson B. and Cornish, Neil J.",
    title = "{Prototype global analysis of LISA data with multiple source types}",
    eprint = "2301.03673",
    archivePrefix = "arXiv",
    primaryClass = "gr-qc",
    doi = "10.1103/PhysRevD.107.063004",
    journal = "Phys. Rev. D",
    volume = "107",
    number = "6",
    pages = "063004",
    year = "2023"
}

@article{Bose:2021pcw,
    author = "Bose, Nirban and Pai, Archana",
    title = "{Effective chirp mass in the inspiral frequency evolution of the nonspinning eccentric compact binary}",
    eprint = "2107.14736",
    archivePrefix = "arXiv",
    primaryClass = "gr-qc",
    reportNumber = "LIGO-P2100280",
    doi = "10.1103/PhysRevD.104.124021",
    journal = "Phys. Rev. D",
    volume = "104",
    number = "12",
    pages = "124021",
    year = "2021"
}

@article{Baird:2012cu,
    author = "Baird, Emily and Fairhurst, Stephen and Hannam, Mark and Murphy, Patricia",
    title = "{Degeneracy between mass and spin in black-hole-binary waveforms}",
    eprint = "1211.0546",
    archivePrefix = "arXiv",
    primaryClass = "gr-qc",
    doi = "10.1103/PhysRevD.87.024035",
    journal = "Phys. Rev. D",
    volume = "87",
    number = "2",
    pages = "024035",
    year = "2013"
}

@misc{nessai,
  author       = {Michael J. Williams and
                  John Veitch and
                  Christian Chapman-Bird},
  title        = {mj-will/nessai: v0.15.2},
  month        = feb,
  year         = 2026,
  publisher    = {Zenodo},
  version      = {v0.15.2},
  doi          = {10.5281/zenodo.18711786},
  url          = {https://doi.org/10.5281/zenodo.18711786},
  swhid        = {swh:1:dir:2a713bd676ead75addbcfd0f0f590da2803853c1
                   ;origin=https://doi.org/10.5281/zenodo.4550693;vis
                   it=swh:1:snp:e58a4ab286164f9416bf15986aebfcda05c77
                   5da;anchor=swh:1:rel:07495132417af7046b7b585d2c75b
                   8148a6bf3a6;path=mj-will-nessai-48ef8bd
                  },
}

@ARTICLE{2024arXiv240812057S,
       author = {{Syed}, Saifuddin and {Bouchard-C{\^o}t{\'e}}, Alexandre and {Chern}, Kevin and {Doucet}, Arnaud},
        title = "{Optimised Annealed Sequential Monte Carlo Samplers}",
      journal = {arXiv e-prints},
         year = 2024,
        month = aug,
          eid = {arXiv:2408.12057},
        pages = {arXiv:2408.12057},
          doi = {10.48550/arXiv.2408.12057},
archivePrefix = {arXiv},
       eprint = {2408.12057},
 primaryClass = {stat.CO},
       adsurl = {https://ui.adsabs.harvard.edu/abs/2024arXiv240812057S}
}

@ARTICLE{1998physics...3008N,
       author = {{Neal}, Radford M.},
        title = "{Annealed Importance Sampling}",
      journal = {arXiv e-prints},
         year = 1998,
        month = mar,
          eid = {physics/9803008},
        pages = {physics/9803008},
          doi = {10.48550/arXiv.physics/9803008},
archivePrefix = {arXiv},
       eprint = {physics/9803008},
 primaryClass = {physics.comp-ph},
       adsurl = {https://ui.adsabs.harvard.edu/abs/1998physics...3008N}
}

@article{Gelman:1992zz,
    author = "Gelman, Andrew and Rubin, Donald B.",
    title = "{Inference from Iterative Simulation Using Multiple Sequences}",
    doi = "10.1214/ss/1177011136",
    journal = "Statist. Sci.",
    volume = "7",
    pages = "457--472",
    year = "1992"
}



\FloatBarrier
\newpage
\onecolumngrid
\section{Supplemental Material}
\section{Generalized SMC with flow-based initialization}\label{sec:smc-details}

In this section, we provide additional mathematical details of the generalized \gls{smc} scheme used in \gls{aspire}, including the construction of the intermediate targets, the incremental importance weights, and the evidence estimator.

\subsection{Intermediate targets and notation}

Let $M_2$ denote the target model with prior $p(\theta|M_2)$, likelihood $p(d|\theta,M_2)$, and posterior
\begin{equation}
    \pi(\theta)
    \equiv p(\theta|d,M_2)
    = \frac{1}{Z}\,p(d|\theta,M_2)\,p(\theta|M_2),
\end{equation}
where $Z \equiv p(d|M_2)$ is the Bayesian evidence. Given an initial analysis under model $M_1$, we train a normalizing flow to approximate the corresponding posterior,
\begin{equation}
    q_\phi(\theta) \approx p(\theta|d,M_1),
\end{equation}
as described in the main text. This density $q_\phi(\theta)$ is used as the initial distribution in our generalized \gls{smc} scheme.

Following Eq.~(2) in the main text, we define a sequence of intermediate targets indexed by an inverse temperature $\beta_t \in [0,1]$,
\begin{equation}
    \pi_t(\theta)
    \equiv p_t(\theta|d,M_2,\beta_t)
    \propto q_\phi(\theta)^{1-\beta_t}
    \bigl[p(d|\theta,M_2)\,p(\theta|M_2)\bigr]^{\beta_t},
    \label{eq:smc:pi_t}
\end{equation}
with $0=\beta_0 < \beta_1 < \cdots < \beta_T = 1$.
By construction,
\begin{equation}
    \pi_0(\theta) = q_\phi(\theta), \qquad
    \pi_T(\theta) \propto p(d|\theta,M_2)\,p(\theta|M_2) = \pi(\theta),
\end{equation}
so that the sequence smoothly interpolates between the flow-based approximation and the desired posterior under $M_2$.

\subsection{Incremental importance weights}

Assume that at iteration $t-1$ we have a set of samples
$\{\theta_i^{(t-1)}, w_i^{(t-1)}\}_{i=1}^N$
with normalized weights $w_i^{(t-1)}$ approximating $\pi_{t-1}$. When moving to $\beta_t$, the unnormalized incremental weight for particle $\theta_i^{(t-1)}$ is given by the standard \gls{smc} update
\begin{equation}
    \tilde{w}_i^{(t)}
    = w_i^{(t-1)}
      \frac{\pi_t\bigl(\theta_i^{(t-1)}\bigr)}
           {\pi_{t-1}\bigl(\theta_i^{(t-1)}\bigr)}.
    \label{eq:smc:incremental_weight_general}
\end{equation}
Using the specific form in Eq.~\eqref{eq:smc:pi_t}, we obtain
\begin{equation}
\begin{split}
    \frac{\pi_t(\theta)}{\pi_{t-1}(\theta)}
    &\propto
    \frac{
        q_\phi(\theta)^{1-\beta_t}
        \bigl[p(d|\theta,M_2)\,p(\theta|M_2)\bigr]^{\beta_t}
    }{
        q_\phi(\theta)^{1-\beta_{t-1}}
        \bigl[p(d|\theta,M_2)\,p(\theta|M_2)\bigr]^{\beta_{t-1}}
    }
    \nonumber \\
    &= 
    q_\phi(\theta)^{-(\beta_t - \beta_{t-1})}
    \bigl[p(d|\theta,M_2)\,p(\theta|M_2)\bigr]^{\beta_t - \beta_{t-1}}.
\end{split}
\end{equation}
It is convenient to define the tempered likelihood–prior ratio
\begin{equation}
    R(\theta)
    \equiv
    \frac{p(d|\theta,M_2)\,p(\theta|M_2)}{q_\phi(\theta)}.
    \label{eq:smc:Rtheta}
\end{equation}
Then the incremental weight simplifies to
\begin{equation}
    \frac{\pi_t(\theta)}{\pi_{t-1}(\theta)}
    \propto R(\theta)^{\Delta\beta_t},
    \qquad
    \Delta\beta_t \equiv \beta_t - \beta_{t-1},
\end{equation}
and the unnormalized weights in Eq.~\eqref{eq:smc:incremental_weight_general} reduce to
\begin{equation}
    \tilde{w}_i^{(t)}
    = w_i^{(t-1)}
      R\bigl(\theta_i^{(t-1)}\bigr)^{\Delta\beta_t}.
    \label{eq:smc:incremental_weight_R}
\end{equation}
The normalized weights are then
\begin{equation}
    w_i^{(t)}
    = \frac{\tilde{w}_i^{(t)}}{\sum_{j=1}^N \tilde{w}_j^{(t)}}.
\end{equation}

In \gls{aspire}, the temperature $\beta_t$ is selected adaptively at each iteration such that the \gls{ess} after reweighting remains close to a target fraction of $N$; we follow the strategy from Ref.~\cite{Karamanis:2022ksp}; see \textit{Adaptive temperature selection and ESS criterion} for more details. Once the weights are updated, we perform resampling followed by a Markov move using a $t$-preconditioned Crank–Nicolson kernel as described in \textit{Particle Resampling and Markov Kernel}, after which the particles are all assigned equal weights $w_{i}^{(t)} = 1 / N$.

\subsection{Evidence estimation and unbiasedness}

The Bayesian evidence for $M_2$ can be expressed as
\begin{equation}
    Z = \int p(d|\theta,M_2)\,p(\theta|M_2)\,\mathrm{d}\theta.
\end{equation}
Let $Z_t$ denote the (unknown) normalizing constant of $\pi_t$, such that
\begin{equation}
    \pi_t(\theta) = \frac{1}{Z_t}\,
    q_\phi(\theta)^{1-\beta_t}
    \bigl[p(d|\theta,M_2)\,p(\theta|M_2)\bigr]^{\beta_t}.
\end{equation}
By construction $Z_0 = \int q_\phi(\theta)\,\mathrm{d}\theta = 1$, while $Z_T = Z$ is the desired evidence. The evidence can be written as a product
\begin{equation}
    Z = Z_T = Z_0 \prod_{t=1}^T \frac{Z_t}{Z_{t-1}}
    = \prod_{t=1}^T \frac{Z_t}{Z_{t-1}}.
    \label{eq:smc:Z_product}
\end{equation}
Each ratio $r_t \equiv Z_t/Z_{t-1}$ can be expressed as an expectation with respect to $\pi_{t-1}$:
\begin{equation}
\begin{split}
    r_t \equiv \frac{Z_t}{Z_{t-1}}
    &= \int \frac{\pi_t(\theta)}{\pi_{t-1}(\theta)}
           \,\pi_{t-1}(\theta)\,\mathrm{d}\theta
     = \mathbb{E}_{\pi_{t-1}}\!\left[
           \frac{\pi_t(\theta)}{\pi_{t-1}(\theta)}
       \right]
    \nonumber \\
    &\propto 
    \mathbb{E}_{\pi_{t-1}}\!\left[
        R(\theta)^{\Delta\beta_t}
    \right],
\end{split}
\end{equation}
with $R(\theta)$ defined in Eq.~\eqref{eq:smc:Rtheta}. Given samples $\{\theta_i^{(t-1)}, w_i^{(t-1)}\}$ approximating $\pi_{t-1}$, we estimate this expectation by
\begin{equation}
    \hat{r_t} \equiv\widehat{\frac{Z_t}{Z_{t-1}}}
    = \sum_{i=1}^N w_i^{(t-1)}\,
      \frac{\pi_t\bigl(\theta_i^{(t-1)}\bigr)}
           {\pi_{t-1}\bigl(\theta_i^{(t-1)}\bigr)}
    = \sum_{i=1}^N \tilde{w}_i^{(t)},
    \label{eq:smc:Z_ratio_est}
\end{equation}
where the last equality follows from Eq.~\eqref{eq:smc:incremental_weight_general} and the normalization of $w_i^{(t-1)}$.
Combining \cref{eq:smc:Z_product,eq:smc:Z_ratio_est}, the \gls{smc} evidence estimator is
\begin{equation}
    \widehat{Z}
    = \prod_{t=1}^T \hat{r_t},
    \qquad
    \log \widehat{Z}
    = \sum_{t=1}^T \log \hat{r_t}.
    \label{eq:smc:Z_estimator}
\end{equation}

We estimate the uncertainty on \cref{eq:smc:Z_estimator} using a delta-method approximation for the variance at each iteration~\cite{Vaart_1998}:
\begin{equation}
    \mathrm{Var}[\widehat r_t]
= \frac{1}{N}\,
  \mathrm{Var}_{\pi_{t-1}}\!\left[
      R(\theta)^{\Delta\beta_t}
  \right].
\end{equation}
Applying a first–order delta method to
$\log \widehat r_t$ gives
\begin{equation}
\mathrm{Var}\!\left[\log \widehat r_t\right]
\approx
\frac{1}{N}\,
\frac{
    \mathrm{Var}_{\pi_{t-1}}\!\left[
        R(\theta)^{\Delta\beta_t}
    \right]
}{
    \mathbb{E}_{\pi_{t-1}}\!\left[
        R(\theta)^{\Delta\beta_t}
    \right]^2
}.   
\end{equation}
Given samples $\{\theta_i^{(t)}, w_i^{(t)}\}$
\begin{equation}
    \widehat{\mathrm{Var}}\!\left[\log \widehat r_t\right]
=
\frac{1}{N}\,
\frac{
    \widehat{\mathrm{Var}}\!\left[\tilde w_i^{(t)}\right]
}{
    \widehat{\mathbb{E}}\!\left[\tilde w_i^{(t)}\right]^2
}.
\end{equation}
If correlations between iterations are neglected, the variances on the log evidence can then be approximated as
\begin{equation}
    \widehat{\mathrm{Var}}[\log \widehat Z]
\approx
\sum_{t=1}^T
\widehat{\mathrm{Var}}\!\left[\log \widehat r_t\right],
\end{equation}
and the standard error reported in this work is
\begin{equation}\label{eq:smc:standard_error}
\widehat{\mathrm{SE}}[\log \widehat Z]
=
\sqrt{
\sum_{t=1}^T
\widehat{\mathrm{Var}}\!\left[\log \widehat r_t\right]
}.
\end{equation}

Under exact sampling from each $\pi_{t-1}$, and in the absence of resampling and Markov moves, \cref{eq:smc:Z_estimator} is an unbiased estimator of $Z$ by standard importance sampling arguments. In practice, \gls{smc} combines reweighting, resampling, and Markov moves; in this setting, the estimator remains consistent and asymptotically unbiased as $N \to \infty$ (see, e.g., \cite{del2012adaptive,Fearnhead:2013}). Crucially, the presence of the flow-based density $q_\phi$ only affects the choice of the initial distribution $\pi_0$ and the intermediate targets $\pi_t$; it does not alter the form of the weights or the evidence estimator. As a result, the use of a learned flow as an initial distribution does not introduce any additional bias in the posterior or evidence estimates beyond the usual Monte Carlo error inherent in \gls{smc}.

\subsection{Adaptive temperature selection and ESS criterion}

We control the annealing schedule $\{\beta_t\}_{t=0}^T$ by targeting a
prescribed \gls{ess} after each reweighting step. Given
normalized weights $\{w_i^{(t)}(\beta)\}_{i=1}^N$, we define
\begin{equation}
  \mathrm{ESS}(\beta)
  = \frac{1}{\sum_{i=1}^N \bigl(w_i^{(t)}(\beta)\bigr)^2},
\end{equation}
where $w_i^{(t)}(\beta)$ are the normalized weights that would result
from updating the inverse temperature from $\beta_t$ to $\beta$ as in
Eq.~\eqref{eq:smc:incremental_weight_R}.

Rather than enforcing a single fixed \gls{ess} fraction throughout the run, we
allow the target \gls{ess} to vary across \gls{smc} stages. Specifically, we define a
stage-dependent \gls{ess} fraction
\begin{equation}
  \rho_{\mathrm{ESS}}^{(t)} = \rho_{\min} + \bigl(\rho_{\max} - \rho_{\min}\bigr) \beta_t^{\gamma},
\end{equation}
which interpolates monotonically between two user-specified values
$0 < \rho_{\min} < \rho_{\max} < 1$, with exponent $\gamma > 0$
controlling the shape of the schedule (e.g.\ increasing from
$\rho_{\min}=0.5$ to $\rho_{\max}=0.8$). When $\rho_{\min} = \rho_{\max}$ this reduces to a fixed target fraction. At each iteration we choose
$\beta_{t+1} > \beta_t$ such that
\begin{equation}
  \mathrm{ESS}(\beta_{t+1})
  \approx \rho_{\mathrm{ESS}}^{(t)} N,
\end{equation}
and solve for $\beta_{t+1}$ via bisection on the interval $(\beta_t, 1]$.
We find that this improves sampling efficiency when including the additional parameters required to model calibration uncertainty in real data.

With this choice and $\rho_{\min}=0.5$ to $\rho_{\max}=0.8$, the total number of \gls{smc} stages $T$ typically lies
between $\sim 5$ and $\sim 40$, depending on the mismatch between
$q_\phi$ and the target posterior.

\subsection{Particle resampling and Markov transitions}
\label{sec:resampling_mcmc}

After each reweighting step in the \gls{smc} procedure we apply multinomial resampling~\cite{douc2005comparison} such that the resulting samples all have equal weights $w_{i}^{(t)} = 1 / N$.

Following resampling, particle diversity is restored through Markov transitions targeting the intermediate distribution $\pi_t$.  
We use a $t$-preconditioned Crank–Nicolson ($t$-pCN) kernel described in Ref.~\cite{2024InvPr..40l5023G} and implemented in the lightweight \texttt{minipcn} package~\cite{minipcn}.  This uses a Student-$t$ distribution for the proposal, the parameters of which are estimated at the beginning of each \gls{smc} step as described in Ref.~\cite{2024InvPr..40l5023G}. Following Ref.~\cite{Karamanis:2025persistent}, we tune the step size during sampling. We apply $n_{\mathrm{steps}}$ \gls{mcmc} steps per \gls{smc} stage.  
For simulated data we use fewer steps, whereas for events where we expect waveform systematics, we adopt a larger value.
The settings are described in more detail \textit{Data generation and inference configuration}.

Once $\beta=1$, we optionally resample the final samples to obtain a user-specified number of samples which are then evolved using the same \gls{mcmc} steps. This yields a predefined number of posterior samples. 

\subsection{Handling parameter mismatches and new parameters}
\label{sec:param_mismatch}

Let $\theta^{(1)} \in \mathbb{R}^{d_1}$ and $\theta^{(2)} \in \mathbb{R}^{d_2}$
denote the parameter vectors for models $M_1$ and $M_2$, respectively,
with $d_2 \ge d_1$. In many of our use cases (e.g.\ adding spin
precession or eccentricity) there is a shared subset of parameters
$\vartheta$ and a set of model-specific parameters
$\psi^{(1)}$ and $\psi^{(2)}$ such that
\[
  \theta^{(1)} = (\vartheta, \psi^{(1)}),
  \qquad
  \theta^{(2)} = (\vartheta, \psi^{(2)}).
\]
Given posterior samples $\{\theta^{(1)}_i\}$ from $M_1$, we construct
initial samples for $M_2$ via
\begin{equation}
  \theta^{(2)}_i = \bigl(\vartheta_i, \tilde{\psi}^{(2)}_i\bigr),
\end{equation}
where $\tilde{\psi}^{(2)}_i \sim p(\psi^{(2)}|M_2)$ are drawn
independently from the prior under $M_2$. In this way, the empirical
distribution of the initial samples approximates
\begin{equation}
  q_\phi(\vartheta, \psi^{(2)}) \approx
  p(\vartheta|d,M_1)\,p(\psi^{(2)}|M_2),
\end{equation}
which matches the desired factorization of the initial density in
\cref{eq:smc:pi_t}. For strongly mismatched dimensions where, for example, known degeneracies between parameters mean the proposal would be uninformative (e.g. adding eccentricity~\cite{Bose:2021pcw}), one can also replace selected components of $\vartheta$ with draws from the prior. A possible criterion for this is discussed in more detail in \textit{Regime of applicability} and the exact choices made for the analyses presented are described in \textit{Analysis details}.

This construction guarantees that the marginal distribution in the new
coordinates is consistent with the prior for the additional parameters,
while still leveraging information from the original posterior in the
shared subspace.

\subsection{Pseudocode for \texorpdfstring{\gls{aspire}}{ASPIRE}}
\label{sec:aspire-pseudocode}

In \cref{alg:aspire-2e}, we summarize the full \gls{aspire} procedure as a flow-initialized generalized \gls{smc} algorithm targeting the posterior under model $M_2$.

\begin{algorithm}[t]
\caption{\gls{aspire}: Flow-initialized Sequential Monte Carlo}
\label{alg:aspire-2e}

\KwIn{Initial posterior samples $\{\theta_i^{(0)}\}$; prior $p(\theta|M_2)$;
likelihood $p(d|\theta,M_2)$; particle count $N$; (optional) desired number of
posterior samples $K \geq N$; minimum and maximum \gls{ess} fractions
$\rho_{\min}, \rho_{\max}$; power $\gamma$; number of MCMC steps
$n_\textrm{steps}$}

\KwOut{Posterior samples for $M_2$ and evidence estimate $\widehat{Z}$.}

Train normalizing flow $q_\phi(\theta)$ on $\{\theta_i^{(0)}\}$\;
Map samples into the $M_2$ parameter space; replace incompatible components with draws from the prior\;

Initialize samples: $\theta_i^{(0)} \sim q_\phi$, $w_i^{(0)}=1/N$,
$\beta_0=0$, $\log\widehat{Z}=0$\;

\While{$\beta_t < 1$}{

    Compute stage-dependent ESS fraction:
    \[
      \rho_{\mathrm{ESS}}
      \leftarrow \rho_{\min}
      + \bigl(\rho_{\max} - \rho_{\min}\bigr)\,\beta_t^{\gamma}
    \]

    Find $\beta_{t+1} > \beta_t$ such that
    $\mathrm{ESS} \approx \rho_{\mathrm{ESS}} N$\;

    Compute incremental weights:
    \[
        \tilde{w}_i^{(t+1)}
        = w_i^{(t)} R(\theta_i^{(t)})^{\Delta\beta_t},
        \quad
        R(\theta)=
        \frac{p(d|\theta,M_2)p(\theta|M_2)}{q_\phi(\theta)}\;
    \]

    Update evidence estimate:\\
    $\log\widehat{Z} \leftarrow \log\widehat{Z}
      + \log\sum_i \tilde{w}_i^{(t+1)}$\;

    Normalize weights:\\
    $w_i^{(t+1)} = \tilde{w}_i^{(t+1)}/\sum_j \tilde{w}_j^{(t+1)}$\;

    Resample samples; set $w_i^{(t+1)} = 1/N$\;

    Run $t$-pCN MCMC targeting $\pi_{t+1}$ for $n_\textrm{steps}$\;

    $t \leftarrow t + 1$\;
}

\If{$K > N$}{
    Resample $\{\theta_i^{(T)}\}_{i=1}^N$ with weights $w_i^{(T)}$ to obtain
    $\{\theta_i^{(T)}\}_{i=1}^K$; set $w_i^{(T)} = 1/K$\;

    Run $t$-pCN MCMC targeting $\pi_{T}$ for $n_\textrm{steps}$ starting from
    $\{\theta_i^{(T)}\}_{i=1}^K$\;
}

\Return{$\{\theta_i^{(T)},w_i^{(T)}\}$ and $\widehat{Z}$}\;

\end{algorithm}

\FloatBarrier
\section{Implementation details}
\phantomsection
\label{sec:implementation-details}

In this section, we describe the practical implementation of \gls{aspire}, covering
the normalizing-flow density estimator and
the resampling scheme and Markov kernel used in the generalized \gls{smc} algorithm. 
Together these components form the basis of the open-source \aspire package~\cite{aspire-doi} and its interfaces for gravitational-wave inference.

\subsection{Normalizing flow}
\phantomsection
\label{sec:flow_implementation}

The first stage of \gls{aspire} trains a normalizing flow $q_\phi(\theta)$ on a set of posterior samples from the initial model $M_1$.  
We use \glspl{maf}~\cite{Papamakarios:2017tec} with four transforms, implemented using the
\texttt{zuko} library~\cite{zuko}. However, the implementation also supports \texttt{jax}-based flows via \texttt{flowjax}~\cite{flowjax}.
Training is performed using the Adam optimizer~\cite{Kingma:2014vow} for 500 epochs with an initial learning rate of 0.001, cosine annealing~\cite{Loshchilov:2016ofz}, and a batch size of 500.  
The trained flow provides both density evaluations $q_\phi(\theta)$ and the ability to generate new samples, which serve as the initial samples for the generalized \gls{smc}.

When models $M_1$ and $M_2$ have different parameters (e.g., when introducing precession or eccentricity), we remap the posterior samples into the parameterization of $M_2$ and replace incompatible components with prior draws, as detailed in \textit{Handling parameter mismatches and new parameters}.  
The flow is then trained on this augmented set, ensuring that the learned density has non-negligible support throughout the region of interest for the target model.

\begin{figure}
    \centering
    \includegraphics[width=0.9\linewidth]{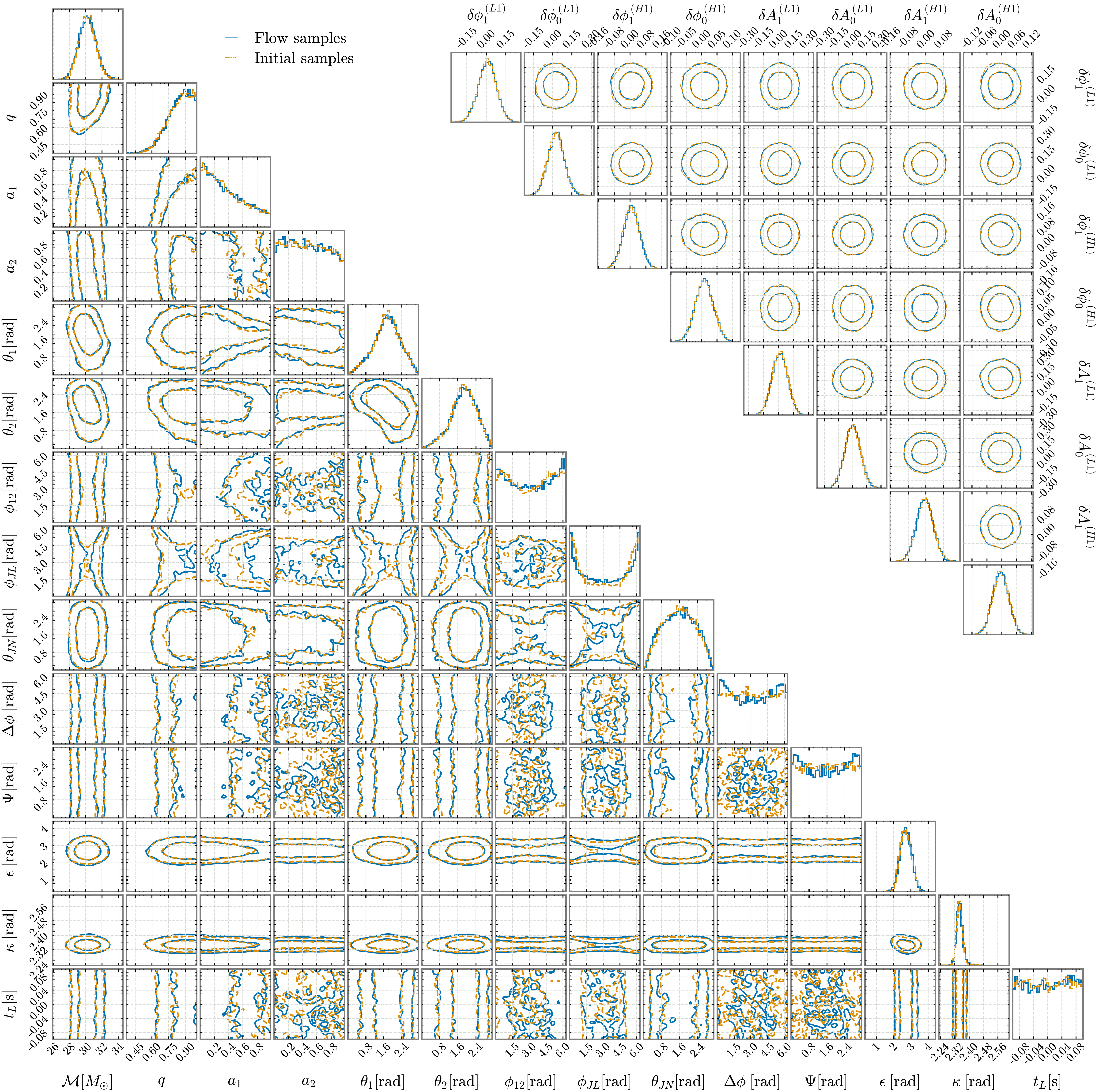}
    \caption{Comparison between the initial samples in orange and distribution learned by the \gls{maf} in blue when analyzing GW150914. The lower plot shows the one- and two-dimensional posterior distributions for the source parameters (excluding luminosity distance, which is marginalized) and the upper panel shows four calibration parameters per detector included in the analysis (H1 and L1). Initial samples were obtained using \dynesty with \texttt{IMRPhenomXPHM}, and the phase, inclination and polarization samples have been drawn from the prior.}
    \label{fig:flow}
\end{figure}

For real-event reanalyses that include calibration parameters, whose posteriors are typically close to Gaussian and only weakly correlated with astrophysical parameters, we adopt a hybrid architecture implemented in \texttt{gwflow}~\cite{gwflow-doi}.  
A multivariate Gaussian (with learned mean and diagonal covariance) is used for the calibration subspace, while a \gls{maf} models the remaining astrophysical parameters.  
We pre-train the Gaussian calibration block for 50 epochs before joint training of the full flow. In \cref{fig:flow}, we show an example of the distribution learned by the flow when trained for GW150914.

\subsection{Software and interfaces}

The complete \gls{aspire} workflow is implemented in the open-source Python package \aspire~\cite{aspire-doi}.  
We also provide \texttt{aspire-bilby} \cite{aspire-bilby-doi}, an interface that enables seamless use of \gls{aspire} within the \bilby inference framework~\cite{Ashton:2018jfp,bilby}, including full support for likelihood, prior, and waveform models used in gravitational-wave astronomy.  
The companion package \texttt{aspire-gw}~\cite{aspire-gw-doi} supports the specialized normalizing-flow architecture described in \textit{Normalizing Flow}.

Together, these tools provide an end-to-end implementation of the flow-initialized \gls{smc} procedure, allow integration with existing Bayesian analysis pipelines, and ensure reproducibility of all results presented in this work. 

\section{Analysis details}

\subsection{Data generation and inference configuration}

All injection studies use simulated signals in a three-detector network at O4 design sensitivity~\cite{KAGRA:2013rdx}. Reference parameter estimation is performed using \bilby with the \dynesty nested sampler. Analyses with \dynesty use the sampler configuration described in Ref.~\cite{LIGOScientific:2025yae}:
\texttt{nlive=1000}, \texttt{naccept=60}, \texttt{checkpoint\_delta\_t=1800}, and
\texttt{sampler=`acceptance-walk'}.
Analyses with the \gls{smc} samples in \aspire use the following settings: \texttt{n\_samples=1000}, \texttt{n\_final\_samples=10000}, \texttt{hidden\_features=[32, 32]}, \texttt{transforms=4} with \texttt{sample\_kwargs} specified as \texttt{sampler="smc"}, \texttt{adaptive=True}, \texttt{target\_efficiency=[0.5, 0.8]}, \texttt{target\_efficiency\_rate=0.5}, \texttt{sampler\_kwargs=dict(n\_steps=500)}. Analyses with real data use a higher target efficiency of: \texttt{[0.5, 0.9]} for GW150914 and \texttt{[0.7, 0.9]} for GW200129. Analyses for asymmetric injection and GW200129 use \texttt{n\_steps=750} and more samples: \texttt{n\_samples=2000} and \texttt{n\_final\_samples=20000} for the asymmetric injection and \texttt{n\_samples=5000} and \texttt{n\_final\_samples=20000} for GW200129. Analyses adding eccentricity use \texttt{n\_steps=300}, and analyses adding precession use \texttt{n\_steps=80}.
All analyses are run using \texttt{bilby\_pipe}. Analyses of simulated data use 16 CPU cores for parallelization, while analyses of real data use 32 CPU cores.

When transitioning between waveform models with incompatible parameterizations, posterior samples of incompatible parameters are replaced by draws from the corresponding prior distributions. This replacement is applied component-wise and preserves correlations among compatible parameters. Specifically, when transitioning between \texttt{IMRPhenomXPHM} and \texttt{IMRPhenomXO4a}, samples of the inclination angle, polarization angle, and coalescence phase are replaced since the waveforms use different conventions for the phase parameter and these three angles are correlated~\cite{Veitch:2014wba}. For analyses introducing spin precession, aligned-spin parameters are replaced by samples drawn from the precessing-spin prior. When introducing eccentricity, both mass and spin parameters are replaced, as there are known degeneracies between eccentricity and chirp mass~\cite{Bose:2021pcw}, and the mass is in turn correlated with the spin parameters~\cite{Baird:2012cu}. In practice, this means that the quasicircular \texttt{TaylorF2} model recovers highly spinning solutions with a bias in chirp mass, see fig. 3 in the Letter.

\subsection{Prior choices and parameterization}

For all analyses, sampling is performed in detector-frame chirp mass $\mathcal{M}$ and mass ratio $q$. Aligned-spin analyses use the aligned components of the spins, while precessing analyses use spin magnitudes and four angles specifying their orientations.

For injections and analyses of real data, an azimuth--zenith sky parameterization is used, with the reference time $t_{\textrm{IFO}}$ defined at the interferometer with the highest signal-to-noise ratio. For probability--probability (P--P) tests, right ascension, declination, and the geocentric reference time are used. In all cases, the luminosity distance is marginalized over~\cite{Thrane:2018qnx}.

Unless otherwise stated, the following priors are used:
\begin{itemize}
\item \textbf{Masses:}
      Sampling is performed in chirp mass $\mathcal{M}$ and mass ratio $q$, with bounds specified in these variables. The prior density includes the appropriate Jacobian such that the induced prior is uniform in the component masses $(m_1, m_2)$.
  \item \textbf{Spins:}
        For aligned-spin analyses, an extension of the prior described in Appendix~A7 of~\cite{Lange:2018pyp}, as implemented in \bilby, is used. For precessing-spin analyses, the spin magnitudes $a_{1,2}$ are drawn uniformly from $[0, \chi_{\max}]$, with spin directions isotropically distributed on the sphere.
  \item \textbf{Luminosity distance:}
        Uniform in comoving volume assuming a Planck15 cosmology.
  \item \textbf{Sky position and orientation:}
        Uniform on the sphere in right ascension $\alpha$ and declination $\delta$ (or azimuth $\epsilon$ and zenith $\kappa$), and uniform in $\cos\theta_{\mathrm{JN}}$ for the inclination angle.
  \item \textbf{Polarization and phase:}
        Uniform in $[0, \pi)$ and $[0, 2\pi)$, respectively.
  \item \textbf{Eccentricity:}
        For \texttt{TaylorF2Ecc}, a prior uniform in
        $e_{20\mathrm{Hz}} \in [0, 0.4]$ is used.
  \item \textbf{Calibration parameters:}
        For each detector and frequency-spline node, Gaussian priors are assumed on amplitude $\delta A_i$ and phase corrections $\delta \phi_i$, with zero mean and standard deviation set by the corresponding calibration uncertainty curves~\cite{Viets:2017yvy}.
\end{itemize}

For the P--P tests, a modified mass prior is adopted following Ref.~\cite{Williams:2021qyt} to avoid oversampling highly asymmetric binaries. Intrinsic masses are drawn from a prior uniform in chirp mass and mass ratio,
\begin{equation}
  p(\mathcal{M}, q) \propto 1,
  \qquad
  \mathcal{M} \in [25, 30]\;\mathrm{M}_\odot,
  \quad
  q \in [0.125, 1],
\end{equation}
and transformed to component masses $(m_1, m_2)$ using standard relations. This prior produces a flat distribution in mass ratio over the range of interest.

\subsection{Details of the probability--probability test}

For the P--P test shown in fig. 5 of the Letter, 100 binary black hole signals are simulated with parameters drawn from the priors described previously. Injection parameters are provided in the accompanying data release~\cite{data_release}. Signals are generated using \texttt{IMRPhenomXPHM} and injected into 8~s of Gaussian noise colored with the O4 design power spectral densities for the three-detector network~\cite{KAGRA:2013rdx}. Each injection is analyzed using \gls{aspire}, initialized from prior samples, and for each parameter the posterior quantile containing the true value is recorded.

All P--P test analyses are performed using a single CPU core.

\subsection{Additional run-time statistics}

In \cref{tab:additional-stats}, we present additional runtime statistics not shown in tab.~1 of the Letter. These show that \gls{aspire} is consistently faster than \dynesty in terms of both wall time and likelihood evaluations.

\begin{table}
    \caption{Run statistics for the analyses presented in the Letter. For each analysis, we quote the number of posterior samples, number of likelihood evaluations and total wall time in hours. \textit{Initial samples} indicates the samples that were used to initialize the analyses where applicable. For \gls{aspire}, likelihood evaluations and wall times are reported as `initial \gls{smc} evolution / total cost', where the latter includes final posterior resampling.}
    \label{tab:additional-stats}
\begin{tabular}{lllC{2.5cm}C{2cm}C{2.5cm}C{2.5cm}}
\toprule
Analysis & Waveform & Sampler & Initial samples & Posterior samples & Likelihood evaluations [$10^6$] & Wall time [hours] \\
\midrule
GW150914-like injection & \texttt{IMRPhenomXPHM} & \texttt{dynesty} & - & 5702 & 30.8 & 12.91 \\
GW150914-like injection & \texttt{IMRPhenomXO4a} & \texttt{dynesty} & - & 5810 & 32.4 & 14.62 \\
GW150914-like injection & \texttt{IMRPhenomXO4a} & \texttt{aspire} & \texttt{IMRPhenomXPHM} + \texttt{dynesty}  & 10000 & 3.03 / 8.03 & 1.53 / 4.01 \\
Asymmetric injection ($q=4$) & \texttt{IMRPhenomXPHM} & \texttt{dynesty} & - & 6259 & 57.9 & 18.01 \\
Asymmetric injection ($q=4$) & \texttt{IMRPhenomXO4a} & \texttt{dynesty} & - & 6511 & 49.6 & 15.68 \\
Asymmetric injection ($q=4$) & \texttt{IMRPhenomXO4a} & \texttt{aspire} & \texttt{IMRPhenomXPHM} + \texttt{dynesty}  & 20000 & 25.6 / 40.6 & 10.29 / 16.02 \\
GW150914-like injection & \texttt{IMRPhenomD} & \texttt{dynesty} & - & 4415 & 7.9 & 1.19 \\
GW150914-like injection & \texttt{IMRPhenomPv2} & \texttt{dynesty} & - & 4737 & 7.15 & 1.16 \\
GW150914-like injection & \texttt{IMRPhenomPv2} & \texttt{aspire} & \texttt{IMRPhenomD} + \texttt{dynesty}  & 10000 & 0.431 / 1.23 & 0.11 / 0.30 \\
Eccentric injection & \texttt{TaylorF2} & \texttt{dynesty} & - & 4564 & 25 & 0.54 \\
Eccentric injection & \texttt{TaylorF2Ecc} & \texttt{dynesty} & - & 4240 & 27.1 & 0.66 \\
Eccentric injection & \texttt{TaylorF2Ecc} & \texttt{aspire} & \texttt{TaylorF2} + \texttt{dynesty}  & 10000 & 4.55 / 7.55 & 0.19 / 0.30 \\
GW150914 (real data) & \texttt{IMRPhenomXPHM} & \texttt{dynesty} & - & 5443 & 34.7 & 9.09 \\
GW150914 (real data) & \texttt{IMRPhenomXO4a} & \texttt{dynesty} & - & 5627 & 32.2 & 11.59 \\
GW150914 (real data) & \texttt{IMRPhenomXO4a} & \texttt{aspire} & \texttt{IMRPhenomXPHM} + \texttt{dynesty}  & 10000 & 4.54 / 9.54 & 1.05 / 2.09 \\
GW200129 (real data) & \texttt{IMRPhenomXPHM} & \texttt{dynesty} & - & 8393 & 65.8 & 12.91 \\
GW200129 (real data) & \texttt{IMRPhenomXO4a} & \texttt{dynesty} & - & 7288 & 53.6 & 19.45 \\
GW200129 (real data) & \texttt{IMRPhenomXO4a} & \texttt{aspire} & \texttt{IMRPhenomXPHM} + \texttt{dynesty}  & 20000 & 37.7 / 52.7 & 11.24 / 13.93 \\
\bottomrule
\end{tabular}

    \centering

\end{table}

\section{Regime of applicability}

In this section, we empirically explore when it is advantageous to use \gls{aspire} and when it would be more efficient to perform a full analysis starting from the prior.

We use two measures of the difference between distributions: the \gls{kld}~\cite{Kullback:1951zyt}:
\begin{equation}
    D_{\textrm{KL}}[p(\theta)||q(\theta)] = \int p(\theta) \log\frac{p(\theta)}{q(\theta)} \textrm{d} \theta
\end{equation}
and the Hellinger distance~\cite{Hellinger1909}:
\begin{equation}
    D_{\textrm{H}^2}[p(\theta)||q(\theta)] = 1 - \int_{} \sqrt{p(\theta)q(\theta)}\,\textrm{d} \theta,
\end{equation}
which is bounded on $[0, 1]$. The latter is particularly relevant for \gls{smc} since it relates to the product of the proposal and posterior distribution, similarly to \cref{eq:smc:pi_t}.

We consider an example in which the prior and likelihood are multivariate Gaussian distributions with diagonal covariance matrices: the prior is $\mathcal{N}({\theta}; {\mu}_0, {\Sigma}_0)$ with ${\mu}_0 = \mathbf{0}$ and $\Sigma_0 = 3 \mathbb{I}$ and the likelihood is $\mathcal{N}({\theta}; {\mu}, \Sigma)$ with ${\mu} = 2\mathbf{1}$ and ${\Sigma} = \mathbb{I}$, where $\mathbf{1}$ denotes a vector of ones. As a result, the posterior is also Gaussian:
\begin{equation}
    p(\theta | d) = \mathcal{N}(\theta;\, \mu_\mathrm{post},\, \Sigma_\mathrm{post}),
\end{equation}
with
\begin{align}\label{eq:gaussian_posterior}
    \Sigma_\mathrm{post} &= \left(\Sigma^{-1} + \Sigma_0^{-1}\right)^{-1}, \\
    {\mu}_\mathrm{post}    &= {\Sigma}_\mathrm{post}\left({\Sigma}^{-1}{\mu} + {\Sigma}_0^{-1}{\mu}_0\right),
\end{align}
We also choose the proposal distribution to be Gaussian $q(\theta) \equiv \mathcal{N}(\theta;\, \mu_q, \Sigma_q)$, where $\Sigma_q = \sigma_q \mathbb{I}$, rather than a normalizing flow, since this allows both measures to be computed analytically.

\begin{figure}
    \centering
    \includegraphics[width=\linewidth]{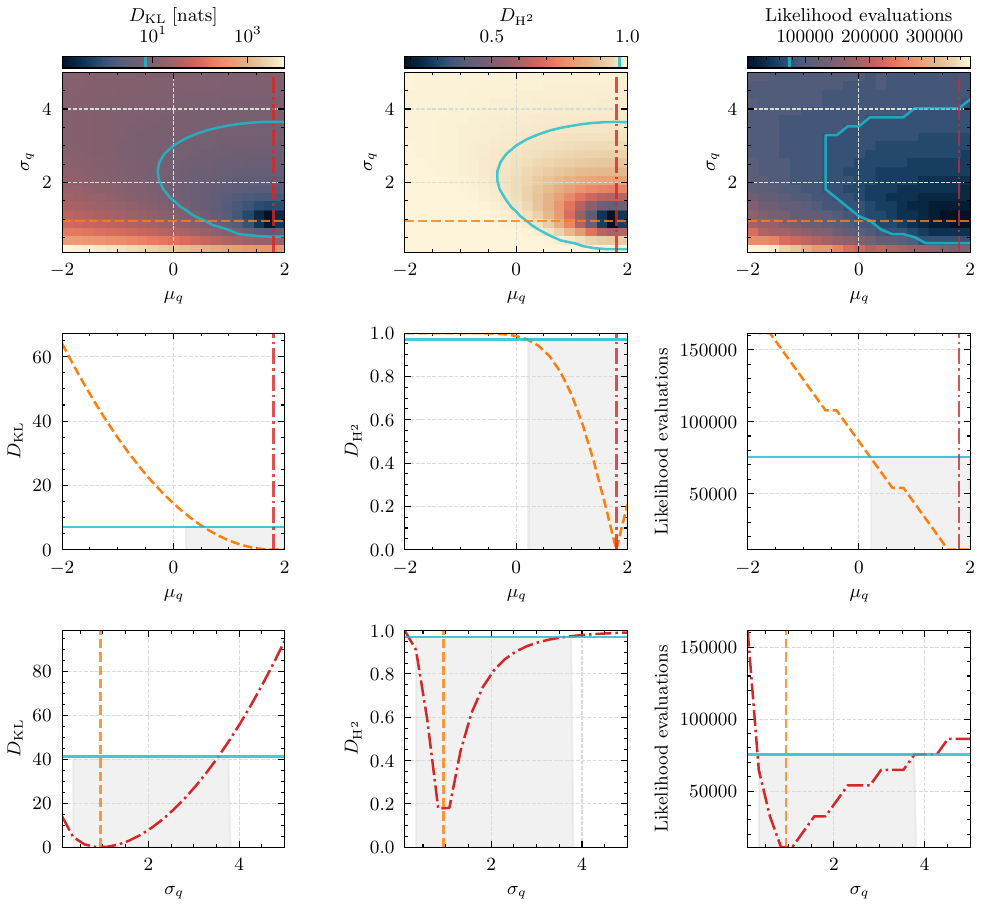}
    \caption{Distance measures and number of likelihood evaluations as a function of proposal mean ($\mu_q$) and standard deviation ($\sigma_q$) when applying \gls{aspire} to an 8-dimensional Gaussian posterior distribution. \textbf{Top:} Kullback--Leibler divergence, Hellinger distance, and total number of likelihood evaluations as a function of proposal mean and standard deviation. The cyan contour shows the value obtained for an analysis starting from the prior rather than $q(\theta)$. The red dash-dot and orange dashed lines indicate the true posterior mean and standard deviation, respectively. \textbf{Middle:} Kullback--Leibler divergence, Hellinger distance, and total number of likelihood evaluations as a function of proposal mean $\mu_q$, with $\sigma_q$ fixed to the true posterior standard deviation. The red dash-dot vertical line indicates the true posterior mean. The cyan horizontal line shows the values obtained for an analysis starting from the prior, and the shaded region indicates where \gls{aspire} requires fewer likelihood evaluations than an analysis starting from the prior. \textbf{Bottom:} The same quantities as a function of proposal standard deviation $\sigma_q$, with $\mu_q$ fixed to the true posterior mean. The orange dashed vertical line indicates the true posterior standard deviation. The cyan horizontal line and shaded region are as above.}
    \label{fig:distances}
\end{figure}

We consider an 8-dimensional parameter space and apply the \gls{aspire} algorithm on a grid of proposal distributions with varying means and standard deviations. At each grid point, we compute the distance between the posterior distribution and the proposal distribution $q(\theta)$, and record the number of inverse-temperature ($\beta$) steps required for \gls{aspire} to converge. We then compare these quantities to those of an analysis starting from the prior and present the results in \cref{fig:distances}. The figure shows that the two distances capture different aspects of the discrepancy between distributions. The $\kld$ decreases as the proposal mean approaches the true posterior mean, but continues to grow without bound as $\sigma_q$ increases, even when  the proposal is already broad enough to cover the posterior. The $\hellinger$ distance, by contrast, responds more symmetrically to both parameters: it drops sharply only when  the proposal is close to the posterior in both mean and standard deviation, and saturates  near unity when the distributions are well-separated.

\begin{figure}
    \centering
    \includegraphics[width=0.5\linewidth]{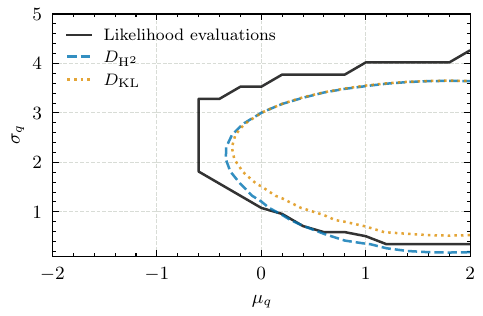}
    \caption{Comparison between the empirical transition boundary where \gls{aspire} becomes cheaper than an analysis starting from the prior (black solid line) and the corresponding contours predicted using the Hellinger distance (blue dashed line) and Kullback--Leibler divergence (orange dotted line).}
    \label{fig:boundaries}
\end{figure}

This difference is most clearly seen in the contour comparison shown in \cref{fig:boundaries}, where both measures approximately track the empirical transition between regimes where \gls{aspire} is and is not advantageous. The Hellinger distance provides a closer approximation to the empirical boundary in some regions of parameter space, particularly for $\sigma_q \lesssim 2$, while the $\kld$ deviates more strongly in this region of parameter space.

These results suggest a simple criterion for applicability: \gls{aspire} is expected to provide a computational advantage when the initial distribution is closer to the target posterior than the prior is. In this regime, the initial samples already capture a significant fraction of the posterior mass, and only a small number of intermediate distributions are required to bridge between them. Conversely, when the initial distribution differs substantially from the target, more $\beta$ steps are required and the computational advantage diminishes.

This can be expressed as an approximate condition for \gls{aspire} to be more efficient than a standard analysis starting from the prior,
\begin{equation}
    \hellinger [p(\theta|d)||q(\theta)] \lesssim \hellinger [p(\theta|d)||p(\theta)].
\end{equation}
That is, the proposal distribution must be closer to the target posterior than the prior is, as measured by the Hellinger distance. In practice, this quantity is not known \emph{a priori}. Future work could explore practical methods for estimating it.

\FloatBarrier
\section{Additional analyses}

In this section, we present additional analyses to further assess the performance and robustness of \gls{aspire}. 
These use toy examples to probe specific failure modes and include comparisons to \gls{ptmcmc}. They are designed to test (i) well-specified initialisation, (ii) biased initialisation, and (iii) multimodal targets.

\subsection{PTMCMC Implementation details}

For a \gls{ptmcmc} sampler, we use \ptemcee~\cite{emcee,Vousden:2016ptemcee}, which has been employed for similar studies in the gravitational-wave literature~\cite{Wolfe:2022nkv} and can be readily interfaced with \aspire via the \texttt{aspire-ptemcee} package, available at \url{https://github.com/mj-will/aspire-ptemcee}. This package includes modifications to support the use of normalizing flows trained within the \aspire framework.

We consider two methods for leveraging the trained normalizing flow within \ptemcee:
\begin{itemize}
    \item \textbf{Seeding chains:} initial samples for each temperature and walker are drawn from the flow. These are the same samples used to initialise the \gls{smc} sampler in \aspire, ensuring a like-for-like comparison.
    \item \textbf{Flow proposal:} the normalizing flow is used as a global proposal distribution across all temperatures and walkers, following the approach employed in \cite{Ashton:2021anp}.
\end{itemize}
There are more sophisticated seeding strategies that one could consider, for example, reweighting nested sampling samples to different temperatures~\cite{Wolfe:2022nkv}. However, since \aspire does not employ such methods (though they could be incorporated), we do not use them here to maintain a fair comparison.

We also ensure that the $\beta = 0$ temperature is included in the temperature ladder. Without this, evidence estimates would be biased because the initial samples are drawn from the normalizing flow rather than the prior.

Log-evidence estimates are computed using two methods: thermodynamic integration and the stepping stone algorithm, both implemented following Ref.~\cite{ANNIS201967}, including their associated variance estimates. For completeness, we also report the coarse error estimate native to \ptemcee, which approximates the uncertainty by recomputing the integral on a thinned temperature grid.

We consider three configurations of \gls{ptmcmc} with \ptemcee, summarised in \cref{tab:ptemcee-settings}. Configuration A follows the settings of Ref.~\cite{Wolfe:2022nkv}; configuration B uses a finer temperature grid; and configuration C replaces the default stretch move with the normalizing flow as the proposal distribution. All three configurations use the default geometric temperature spacing from \ptemcee. During post-processing, we discard the burn-in specified in \cref{tab:ptemcee-settings} and thin the chains according to the mean estimated autocorrelation length. To assess convergence, we compute the Gelman–Rubin $\hat{R}$ statistic~\cite{Gelman:1992zz} for the \gls{mcmc} chains. For the cold chains corresponding to the target posterior ($\beta=1$), unless stated otherwise, all parameters satisfied $\hat{R}<1.01$.

\begin{table}
    \caption{Settings used for \gls{ptmcmc} analyses with \ptemcee. Configuration A follows Ref.~\cite{Wolfe:2022nkv}; B uses a finer temperature grid; C uses the normalizing flow trained during the initial \gls{aspire} step as the proposal distribution.}
    \centering
    \begin{tabular}{lccccc}
        \toprule
        & \texttt{ntemps} & \texttt{nwalkers} & \texttt{nsteps} &  Burn-in &Proposal \\
        \midrule
        PTMCMC A & 5  & 250 & 2000 & 100 & Stretch \\
        PTMCMC B & 16 & 250 & 2000 & 100 & Stretch \\
        PTMCMC C & 16  & 250 & 2000 & 200 & Normalizing flow \\
        \bottomrule
    \end{tabular}
    \label{tab:ptemcee-settings}
\end{table}

\subsection{Toy examples}

We consider three toy examples to probe the robustness of \gls{aspire} in controlled settings and also compare to \gls{ptmcmc}.

\subsubsection{15-dimensional Gaussian}

The first test uses a multivariate Gaussian likelihood in 15 dimensions, matching the dimensionality of the standard \gls{cbc} inference problem. The prior is also chosen to be Gaussian, so that the true posterior and evidence can be computed analytically, and the initial samples are drawn from another Gaussian. This probes the default regime of use for \gls{aspire}, where the initial distribution is
close to the target distribution.

The likelihood is $\mathcal{N}({\theta}; {\mu}, {\Sigma})$ with ${\mu} = 2\mathbf{1}$ and ${\Sigma} = \mathbb{I}$, and the prior is $\mathcal{N}({\theta}; {\mu}_0, {\Sigma}_0)$ with ${\mu}_0 = \mathbf{1}$ and ${\Sigma}_0 = 2\mathbb{I}$, where $\mathbf{1}$ is a vector of ones. The posterior is then also Gaussian given by \cref{eq:gaussian_posterior} and the log-evidence is
\begin{equation}
    \ln \mathcal{Z} = \ln \mathcal{N}\!\left({\mu};\, {\mu}_0,\, {\Sigma} + {\Sigma}_0\right).
\end{equation}
We draw 5,000 initial samples for training the flow from another Gaussian distribution: $\theta \sim \mathcal{N}({\mu_\textrm{Init}}, {\Sigma_\textrm{Init}})$ with ${\mu} = 1.5\mathbf{1}$ and ${\Sigma} = \mathbb{I}$.

We run the \texttt{minipcn}-based \gls{smc} sampler in \aspire with the following settings: \texttt{n\_samples=5000}, \texttt{n\_final\_samples=10000}, \texttt{target\_efficiency=0.9}, \texttt{sampler\_kwargs=dict(n\_steps=50))} and the \ptemcee-based \gls{ptmcmc} sampler using the settings described in \cref{tab:ptemcee-settings}. We use the same trained normalizing flow for all the analyses.

Both methods recover posterior distributions that are consistent with the true distribution. The log-evidence estimates from each method are summarised in \cref{tab:gaussian-evidence}. The result from \gls{aspire} is in excellent agreement with the analytic value.

For the \gls{ptmcmc} configurations, the behaviour depends on the estimator. The stepping stone estimates are generally close to the analytic value, with PTMCMC A in particularly good agreement, while PTMCMC B and C show modest deviations. In contrast, thermodynamic integration yields estimates that are systematically offset from the analytic value, despite reporting small nominal uncertainties. The larger, coarse uncertainty estimates reported by \ptemcee (shown in parentheses) are more consistent with the observed discrepancies and, in some cases, encompass the true value. For all \gls{ptmcmc} configurations, the chains satisfied $\hat{R} < 1.01$ for all parameters and temperatures.

Notably, increasing the number of temperatures (PTMCMC B relative to A) does not improve agreement with the analytic value, indicating that the discrepancies cannot be attributed solely to insufficient temperature resolution. Instead, they likely reflect a combination of suboptimal temperature placement and incomplete mixing, particularly at high temperatures. These results highlight that, while \gls{ptmcmc} is in principle capable of accurate evidence estimation, achieving reliable results in practice requires careful tuning of both the temperature ladder and mixing across temperatures.

\begin{table}\label{eq:toy:biased_post}
    \caption{Log-evidence estimates for the 15-dimensional Gaussian test case. For \gls{ptmcmc}, results are shown using both thermodynamic integration and the stepping stone estimator and their corresponding uncertainties. The coarse uncertainty estimate used in \ptemcee is shown in parentheses.  Log-evidence estimates that are within 3-$\sigma$ of the analytic result are highlighted in bold. For \gls{aspire}, likelihood evaluations are reported as `initial \gls{smc} evolution / total cost', where the latter includes final posterior resampling.}
    \label{tab:gaussian-evidence}
    \begin{tabular}{p{2.5cm} C{2cm} C{3 cm} C{2.5cm} c c}
\toprule
Method & Posterior samples & Likelihood evaluations & Estimator & $\log Z - \log Z_{\mathrm{analytic}}$ & Uncertainty \\
\midrule
ASPIRE SMC & 10,000 & 1,065,000 / 1,565,000 & \Cref{eq:smc:Z_estimator} & \textbf{0.00} & 0.01 \\
\midrule
\multirow{2}{*}{PTMCMC A}
 & 22,750 & 2,502,500 & Thermodynamic Integration & -0.17 & 0.01 (0.64) \\
 &  &  & Stepping Stone & \textbf{0.05} & 0.02 \\
\midrule
\multirow{2}{*}{PTMCMC B}
 & 53,000 & 8,008,000 & Thermodynamic Integration & 0.13 & 0.01 (0.41) \\
 &  &  & Stepping Stone & 0.28 & 0.02 \\
\midrule
\multirow{2}{*}{PTMCMC C}
 & 50,000 & 8,008,000 & Thermodynamic Integration & 0.11 & 0.01 (0.41) \\
 &  &  & Stepping Stone & 0.26 & 0.02 \\
\bottomrule
\end{tabular}

\end{table}

\subsubsection{Biased initial samples}
\label{sec:biased-initial-samples}

The second example emulates a scenario in which the initial distribution used to seed \gls{aspire} has different support from the true posterior. This tests the robustness of each sampler when the initial distribution is deliberately misspecified.

\begin{figure}
    \centering
    \includegraphics[width=\linewidth]{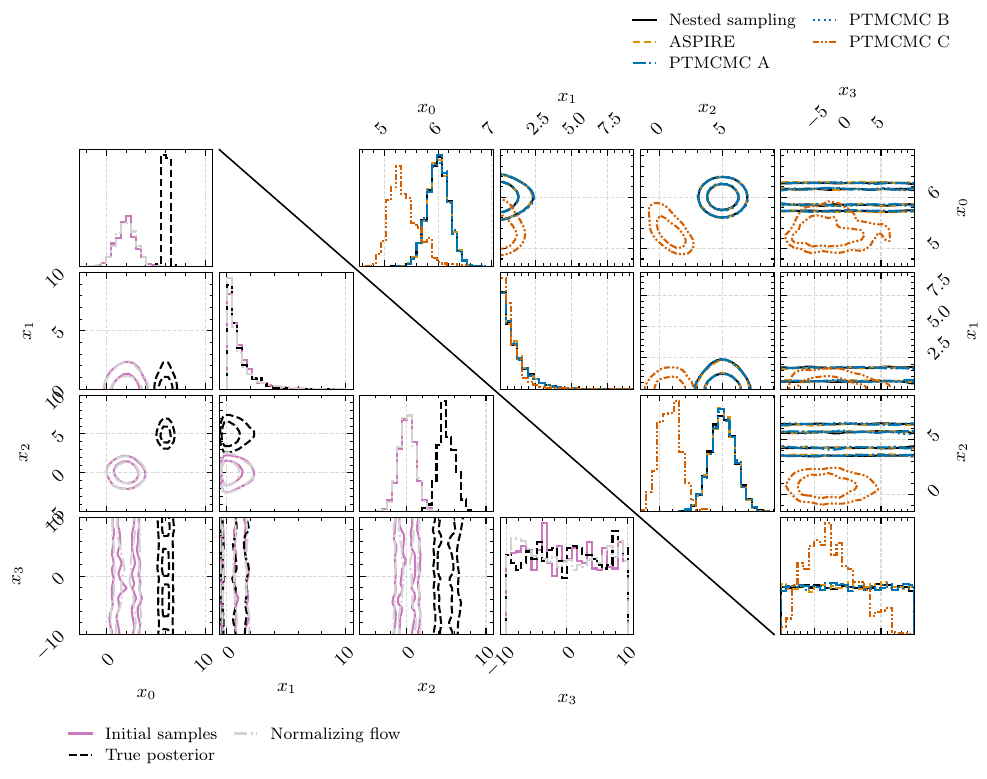}
    \caption{Initial samples and posterior samples for the toy example with biased initial samples. \textbf{Lower triangle:} Initial samples drawn according to \cref{eq:toy:biased_dist}, samples drawn from the trained flow and samples drawn from the true posterior distribution \cref{eq:toy:biased_dist}. \textbf{Upper triangle:} posterior samples obtained with different sampling methods: nested sampling, \gls{smc} and \gls{ptmcmc} with three different configurations (see \cref{tab:ptemcee-settings}).}
    \label{fig:biased_initial_samples}
\end{figure}

We construct a four-dimensional target in which the marginal posteriors are independent:
\begin{align}
    p(\theta_0 \mid {d}) &\propto \mathcal{N}(\theta_0;\, 6,\, 0.2^2), \\
    p(\theta_1 \mid {d}) &\propto \mathrm{Exp}(\theta_1;\, 1), \\
    p(\theta_2 \mid {d}) &\propto \mathcal{N}(\theta_2;\, 5,\, 1), \\
    p(\theta_3 \mid {d}) &\propto \mathrm{Uniform}(\theta_3;\, -10,\, 10),
\end{align}
with uniform priors on all parameters, so that samples from the true posterior can be drawn analytically. The biased initial samples are drawn from a different distribution with mismatched location and scale in parameters $\theta_0$ and $\theta_1$:
\begin{align}\label{eq:toy:biased_dist}
    \theta_0^{(0)} &\sim \mathcal{N}(2,\, 1), \\
    \theta_1^{(0)} &\sim \mathrm{Exp}(1), \\
    \theta_2^{(0)} &\sim \mathcal{N}(0,\, 1), \\
    \theta_3^{(0)} &\sim \mathrm{Uniform}(-10,\, 10).
\end{align}
\Cref{fig:biased_initial_samples} shows the initial samples and true posterior distribution below the main diagonal.

We draw 5,000 initial samples and train a \gls{maf} using default settings with \texttt{hidden\_features=[32, 32]} for 100 epochs. The distribution learnt by the flow is also shown in \cref{fig:biased_initial_samples}. We then use that flow as the proposal for an analysis with the \texttt{minipcn}-based \gls{smc} sampler in \aspire with the following settings: \texttt{n\_samples=1000, n\_final\_samples=10000, target\_efficiency=0.8, sampler\_kwargs=dict(n\_steps=16))} and the \ptemcee-based \gls{ptmcmc} sampler using the settings described in \cref{tab:ptemcee-settings}. For comparison, we also performed a nested sampling analysis starting from the prior using \texttt{nessai}~\cite{Williams:2021qyt,nessai} with \texttt{nlive=5000}, which we treat as the ground truth.

We show the posterior samples produced by each method in \cref{fig:biased_initial_samples}. Both \gls{aspire} and \gls{ptmcmc} are able to recover posterior distributions that are consistent with nested sampling despite the biased initialisation, however, \gls{ptmcmc} performs poorly when using the normalizing flow as a proposal, as the proposal concentrates probability mass in regions that do not adequately cover the posterior. For \gls{ptmcmc} configurations A and B, the chains satisfied $\hat{R} < 1.01$ for all parameters and temperatures, whereas for configuration C, $\hat{R} > 1.01$ for the higher temperature chains confirming the chains have not reached a steady state.

\begin{figure}
    \centering
    \includegraphics[width=\linewidth]{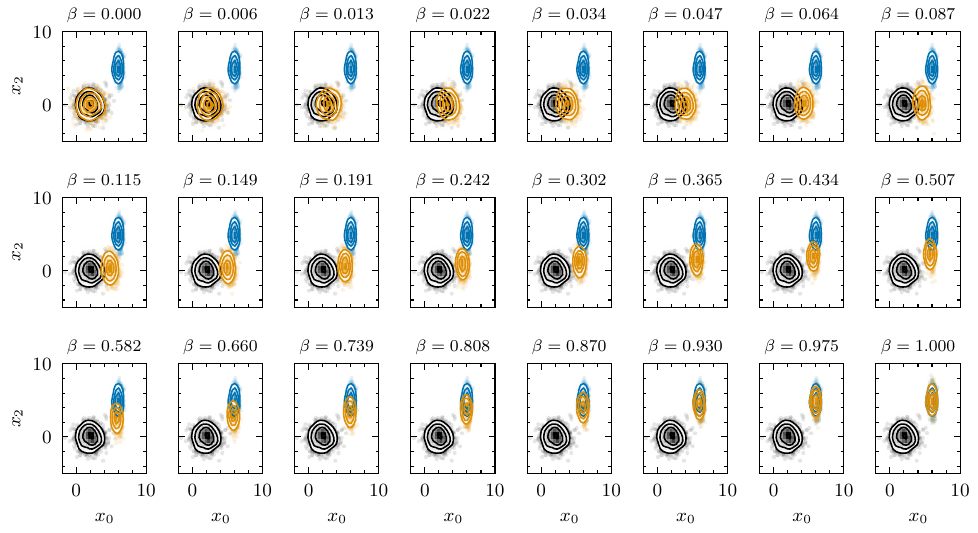}
    \caption{Evolution of intermediate target distributions during an analysis with \aspire. Samples are shown for $\theta_0$ and $\theta_2$ from the \textit{Biased initial samples} toy example. The black samples show the initial samples from $q(x)$ and the blue samples are samples drawn analytically from the posterior distribution. The orange samples show the intermediate samples for a given inverse temperature $\beta$.}
    \label{fig:smc-steps}
\end{figure}

\Cref{fig:smc-steps} illustrates how \gls{aspire} bridges the gap between the biased initial distribution and the true posterior through a sequence of intermediate targets parameterised by the inverse temperature $\beta$. At $\beta=0$, samples follow the initial proposal $q(x)$, which places substantial probability mass away from the posterior support. As $\beta$ increases, the posterior is gradually introduced, and the intermediate distributions smoothly deform towards the true posterior. This annealing process allows the sampler to progressively reweight and move samples into the high-probability regions of the target, even when the initial distribution has poor overlap with the posterior.

The evidence estimates for each method are shown in \cref{tab:biased-evidence}. \gls{aspire}'s \gls{smc}-based estimate is consistent with the estimate from nested sampling. However, the evidence estimates from \gls{ptmcmc} once again show significant variability across configurations and are sensitive to the choice of temperature ladder. The stepping stone algorithm's estimates are consistent for both configurations A and B, however the estimate from thermodynamic integration is not.

\begin{table}
    \caption{Log-evidence estimates for the biased initial samples toy example. For \gls{ptmcmc}, results are shown using both thermodynamic integration and the stepping stone estimator and their corresponding uncertainties. The coarse uncertainty estimate used in \ptemcee is shown in parentheses. Log-evidence differences are shown relative to a reference estimate computed using a conservative nested-sampling run with \texttt{nessai}. Values that are within 3-$\sigma$ of the reference estimate are highlighted in bold. For \gls{aspire}, likelihood evaluations are reported as `initial \gls{smc} evolution / total cost', where the latter includes final posterior resampling.}
    \label{tab:biased-evidence}
    \begin{tabular}{l C{2.5cm} C{2.5cm} C{3.2cm} c c}
\toprule
Method & Posterior samples & Likelihood evaluations & Estimator & $\log Z - \log Z_{\mathrm{reference}}$ & Uncertainty \\
\midrule
Nested Sampling (from prior)
 & 18,649
 & 120,183
 & Skilling \cite{Skilling:2004pqw}
 & \textbf{0.00}
 & 0.03 \\
\midrule
ASPIRE
 & 10,000
 & 363,000 / 523,000
 & \Cref{eq:smc:Z_estimator}
 & \textbf{0.04}
 & 0.07 \\
\midrule
\multirow{2}{*}{PTMCMC A}
 & \multirow{2}{*}{11,250}
 & \multirow{2}{*}{1,861,480}
 & Thermodynamic Integration
 & -9.79
 & 0.11 (28.12) \\
 &  &  & Stepping Stone
 & \textbf{-0.03}
 & 0.03 \\
\cmidrule(lr){1-6}
\multirow{2}{*}{PTMCMC B}
 & \multirow{2}{*}{16,500}
 & \multirow{2}{*}{5,584,414}
 & Thermodynamic Integration
 & -1.39
 & 0.03 (4.75) \\
 &  &  & Stepping Stone
 & \textbf{0.03}
 & 0.04 \\
\cmidrule(lr){1-6}
\multirow{2}{*}{PTMCMC C}
 & \multirow{2}{*}{5,000}
 & \multirow{2}{*}{8,008,000}
 & Thermodynamic Integration
 & -9.89
 & 0.05 (4.65) \\
 &  &  & Stepping Stone
 & -8.37
 & 0.11 \\
\bottomrule
\end{tabular}
\end{table}

\subsubsection{Multimodal target}

The third example considers a multimodal target distribution where the initial proposal only captures a subset of the posterior support. This probes a potential failure mode of flow-based initialisation, namely incomplete coverage of multimodal targets.

We define a simple four-dimensional multimodal target consisting of a mixture of two Gaussians:
\begin{equation}
    p(d|\theta) = \frac{1}{2} \left[\mathcal{N}(\theta; \mu_0, \Sigma_0) + \mathcal{N}(\theta; \mu_1, \Sigma_1) \right],
\end{equation}
where $\mu_0=\mathbf{0}$, $\mu_1=\mathbf{5}$ and $\Sigma_0=\Sigma_1=\mathbb{I}$. We adopt a uniform prior $\textrm{Uniform}(\theta;-10, 10)$. In this case, the evidence can be computed analytically and is given by the inverse prior volume,
\begin{equation}
    \ln Z = -\ln 20^{4}.
\end{equation}

We draw 5,000 initial samples from a Gaussian centred on $\mu_0$, so that only one of the modes is represented in the initial proposal. We run the \texttt{minipcn}-based \gls{smc} sampler in \aspire with \texttt{n\_samples=1000, target\_efficiency=0.9, n\_final\_samples=5000, sampler\_kwargs=dict(n\_steps=200)}, and the \ptemcee-based \gls{ptmcmc} sampler using configuration A from \cref{tab:ptemcee-settings}. The same trained normalizing flow is used in all cases.

\begin{figure}
    \centering
    \includegraphics[width=0.8\linewidth]{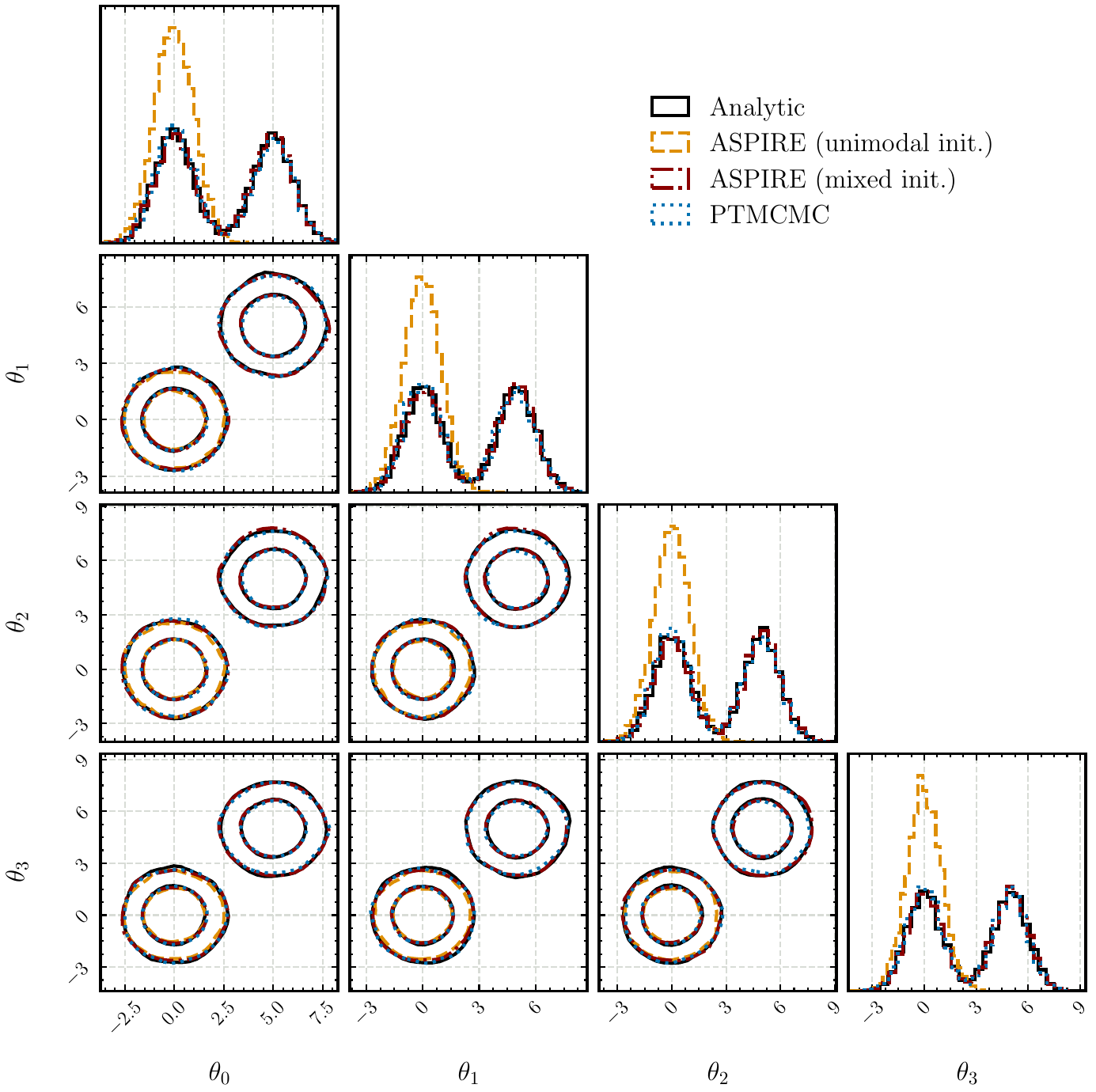}
    \caption{Posterior samples for the multimodal toy example, where the initial distribution is unimodal. Samples drawn from the analytic posterior are shown alongside results from two \gls{aspire} analyses: one initialized from samples covering only a single mode, and the other using a mixture of those samples and draws from the prior. Results obtained with \gls{ptmcmc}, initialised from the same unimodal distribution, are also shown. This highlights that \gls{aspire} fails to recover the second mode when the initial proposal lacks support, but succeeds when even a small fraction of prior samples is included.}
    \label{fig:toy-examples:multi}
\end{figure}

The results are shown in \cref{fig:toy-examples:multi} and \cref{tab:toy-examples:multi}. We find that \gls{aspire} fails to recover the second mode when it is absent from the initial proposal, instead concentrating on the mode present in the initial samples. This reflects a support mismatch: regions of high posterior probability that are not represented in the initial proposal cannot be recovered during the SMC updates. As a result, the log-evidence is underestimated. In contrast, \gls{ptmcmc} successfully explores both modes and recovers an evidence consistent with the analytic value. For both \gls{ptmcmc} configurations, the chains satisfied $\hat{R} < 1.01$ for all parameters and temperatures.

This behaviour reflects a key difference between the methods. Since \gls{ptmcmc} includes a $\beta=0$ chain sampling from the prior, it can in principle discover modes not present in the initialisation. \Gls{aspire}, instead, depends on the support of the initial proposal and may fail when the initial samples occupy only one of several disconnected modes, since the SMC updates can remain concentrated on that mode even when another mode has comparable posterior mass.

To mitigate this failure mode, we repeat the analysis after replacing 20\% of the initial samples with draws from the prior. This modification has little impact on the \gls{ptmcmc} results, as expected, but significantly improves the performance of \gls{aspire}, allowing it to recover both modes, as seen in \cref{fig:toy-examples:multi}. Correspondingly, the log-evidence estimate shifts from a biased value, arising from incomplete mode coverage, to one consistent with the analytic result, as shown in \cref{tab:toy-examples:multi}. This demonstrates that even a small amount of prior support in the initialisation can be sufficient to restore accurate inference.

\begin{table}[]
    \centering
    \caption{Log-evidence estimates and run-time statistics for the multimodal toy example. Estimates within $3\sigma$ of the analytic value are highlighted in bold. Log-evidence values for \gls{ptmcmc} are computed using the stepping stone algorithm. For \gls{aspire}, likelihood evaluations are reported as `initial \gls{smc} evolution / total cost', where the latter includes final posterior resampling.}
    \label{tab:toy-examples:multi}
    \begin{tabular}{p{5cm} C{2.5cm} C{3cm} c c}
\toprule
Method & Posterior samples & Likelihood evaluations & $\log Z - \log Z_{\mathrm{analytic}}$ & Uncertainty \\
\midrule
ASPIRE SMC (unimodal init.) & 10,000 & 223,000 / 2,223,000 & -0.69 & 0.00 \\
ASPIRE SMC (mixed init.) & 10,000 & 2,243,000 / 4,243,000 & \textbf{-0.01} & 0.02 \\
PTMCMC (unimodal init.) & 6,000 & 2,055,287 & \textbf{-0.01} & 0.04 \\
PTMCMC (mixed init.) & 6,000 & 2,054,289 & \textbf{0.01} & 0.04 \\
\bottomrule
\end{tabular}

\end{table}

\subsection{Summary of findings}

These toy examples clarify the regimes in which \gls{aspire} is most effective. When the initial proposal provides adequate coverage of the posterior support, \gls{aspire} achieves comparable or greater accuracy than \gls{ptmcmc} while requiring substantially fewer likelihood evaluations. In this regime, \gls{aspire} provides a computationally efficient approach for updating existing analyses.

When the initial samples are biased relative to the target, both \gls{aspire} and \gls{ptmcmc} recover the correct posterior; however, for the configurations considered here, \gls{ptmcmc} exhibits variability in the evidence estimates, while \gls{aspire} remains stable and significantly more efficient.

In cases where the initial proposal lacks support for a mode of comparable posterior probability, \gls{aspire} can fail to recover that mode. This behaviour arises from the dependence of SMC on the support of the initial distribution. However, we show that this limitation can be mitigated straightforwardly by augmenting the initial samples with a small fraction of prior draws, which restores accurate posterior and evidence estimation.

Overall, these results highlight a trade-off between efficiency and global robustness. In the regimes most relevant for sequential gravitational-wave analyses, where updated models are expected to remain close to previous posteriors, \gls{aspire} provides a more efficient alternative to conventional sampling methods.

\FloatBarrier
\subsection{GW150914-like example with PTMCMC}

We also repeat the GW150914-like injection analysis using \gls{ptmcmc} with \ptemcee. In particular, we consider the case of transitioning between \texttt{IMRPhenomXPHM} and \texttt{IMRPhenomXO4a} and use the \ptemcee B sampling configuration described in \cref{tab:ptemcee-settings}. We do not use the configuration with the normalizing flow proposal, as the toy examples indicate that this proposal can be unreliable for this class of problems. This example provides a direct comparison to the results presented in the Letter.

\begin{figure}
    \centering
    \includegraphics[width=0.7\linewidth]{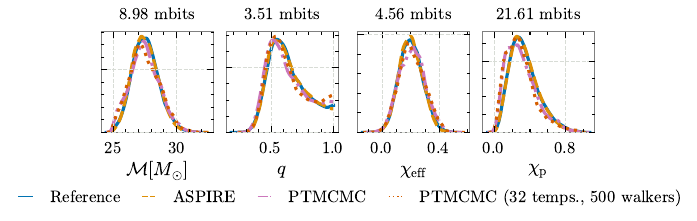}
    \caption{Posterior distributions for a GW150914-like system with comparable masses and low spins when analyzed with \dynesty, \aspire and \ptemcee using \texttt{IMRPhenomXO4a}. The \aspire and \ptemcee analyses are initialized from the analysis shown in fig. 2 of the Letter that uses \dynesty with \texttt{IMRPhenomXPHM}. For \ptemcee we show two analyses, one using configuration B from \cref{tab:ptemcee-settings} and a second where the number of temperatures and walkers are doubled to 32 and 500 respectively. The \gls{jsd} in millibits (mbits) between the \dynesty and \ptemcee results with 32 temperatures is quoted above each subplot.}
    \label{fig:ptmcmc:gw150914-like}
\end{figure}

In \cref{fig:ptmcmc:gw150914-like} we compare the posterior distributions obtained with \gls{ptmcmc} to those present in fig.~2 of the Letter. We observe significant disagreement between the posterior distributions obtained with \gls{ptmcmc} and those from \dynesty and \aspire, suggesting convergence difficulties in the \gls{ptmcmc} chains under these configurations. We repeat the analysis using more stringent settings (\texttt{nwalkers=500}, \texttt{ntemps=32}), but find that this does not significantly improve the agreement, suggesting that the issue is not simply due to insufficient temperature resolution. 
For all configurations, the cold chains corresponding to the target posterior satisfy $\hat{R} < 1.01$ for all parameters, indicating that the posterior chains themselves are well converged. However, several higher-temperature chains exhibit larger $\hat{R}$ values, consistent with poor mixing across temperatures and incomplete exploration of the parameter space. This can lead to inaccurate posterior and evidence estimates in \gls{ptmcmc}, despite apparently well-converged cold chains.

In \cref{tab:ptmcmc:gw150914-evidence}, we present the inferred evidence estimates and run-time statistics for the four analyses. This shows that, in addition to the differences in the inferred posterior distributions, the corresponding evidence estimates also differ significantly from those obtained with \dynesty and \aspire. The wall time and number of likelihood evaluations per sample are comparable to \gls{aspire}, however, since the posterior and evidence estimates are inconsistent with the reference analyses, these metrics alone do not provide a meaningful comparison of performance.

\begin{table}
    \centering
    \caption{Log evidences and run-time statistics for the four analyses presented in \cref{fig:ptmcmc:gw150914-like}. The \dynesty and \aspire results are the same as those presented in the Letter. The table shows the number of posterior samples produced by each method, number of likelihood evaluations, wall time in hours and the corresponding values per posterior sample. For \gls{aspire}, likelihood evaluations and wall times are reported as `initial \gls{smc} evolution / total cost', where the latter includes final posterior resampling.}
    \begin{tabular}{p{3.5cm}C{2cm}cC{2.5cm}C{2.5cm}cC{2.5cm}}
\toprule
Sampler & Posterior Samples & Log Evidence & Likelihood evaluations [$10^6$] & Wall time [hours] \\
\midrule
\texttt{dynesty} & 5810 & -57893.231404 & 32.4 & 14.62 \\
ASPIRE SMC & 10000 & -57893.360252 & 3.03 / 8.03 & 1.53 / 4.01 \\
PTMCMC B & 9500 & -57885.684353 & 4.83 & 2.32 \\
PTMCMC B (32 temps., 500 walkers) & 19000 & -57885.268700 & 19.0 & 7.55 \\
\bottomrule
\end{tabular}

    \label{tab:ptmcmc:gw150914-evidence}
\end{table}

These results indicate that, in these configurations, \gls{ptmcmc} does not provide a direct replacement for the \gls{smc} component of \gls{aspire}. 
Overall, this example highlights that, in realistic high-dimensional problems, \gls{ptmcmc} can exhibit convergence challenges and may require careful tuning of algorithmic parameters beyond simply increasing computational effort. Alternative initialisation or proposal strategies (e.g.~\cite{Wolfe:2022nkv}) may improve performance and could also be incorporated within \gls{aspire}; we leave a detailed exploration of these approaches to future work.

\end{document}